\documentclass{aa}
\usepackage[varg]{txfonts}
\usepackage{graphicx}
\pdfoutput=1

\begin{document}

\title{Chromospheric swirls}
   \subtitle{I. Automated detection in H$\alpha$ observations and their statistical properties}

   \author{I. Dakanalis
          \inst{1,2}
          \and
          G. Tsiropoula\inst{2}\fnmsep
          \and
          K. Tziotziou\inst{2}
          \and
          I. Kontogiannis\inst{3}
          }
   \institute{Department of Physics, National and Kapodistrian University of Athens (UOA)\\
              \email{idakanalis@noa.gr}
         \and
             Institute for Astronomy, Astrophysics, Space Applications and Remote Sensing (IAASARS)\\
             \email{georia@noa.gr, kostas@noa.gr}
        \and
             Leibniz-Institut f\"{u}r Astrophysik Potsdam (AIP)
             }
%context ;aims ;methods ;results ;conclusions
\abstract{Chromospheric swirls are related to convectively driven vortex flows and considered to play a significant role in the dynamics and heating of the upper solar atmosphere. It is important to automatically detect and track them in chromospheric observations and determine their properties.}{We aim to detect and track chromospheric swirls both in space and time by applying a newly developed novel automated method on high quality time series of H$\alpha$ observations and to conduct a statistical analysis to determine their properties.}{We applied a recently developed automated chromospheric swirl detection method to time-series observations of a quiet region of the solar chromosphere obtained in the H$\alpha$-0.2\,{\AA} wavelength of the H$\alpha$ spectral line by the CRISP instrument at the Swedish 1-m Solar Telescope. The algorithm exploits the morphological characteristics of swirling events in high contrast chromospheric observations and results in the detection of these structures in each frame of the time series and their tracking over time. We conducted a statistical analysis to determine their various properties, including a survival analysis for deriving the mean lifetime.}{A mean number of 146\,$\pm$\,9 swirls was detected within the H$\alpha$-0.2\,{\AA} field of view at any given time. The mean surface density is found equal to $\sim$0.08\,swirls$\,$Mm$^{-2}$ and the occurrence rate is $\sim$10$^{-2}$\,swirls$\,$Mm$^{-2}$\,min$^{-1}$. These values are much higher than those previously reported from chromospheric observations. The radii of the detected swirls range between 0.5 and 2.5\,Mm, with a mean value equal to 1.3\,$\pm$\,0.3\,Mm, which is slightly higher than previous reports. The lifetimes range between 1.5\,min and 33.7\,min (equal to the duration of the observations) with an arithmetic mean value of $\sim$8.5\,min. A survival analysis of the lifetimes, however, using the Kaplan-Meier estimator in combination with a parametric model results in a mean lifetime of 10.3\,$\pm$\,0.6\,min.}{Swirls are ubiquitous in the solar chromosphere. An automated method sheds more light on their abundance than visual inspection, while higher cadence, higher resolution observations will most probably result in the detection of a higher number of such features on smaller scales and with shorter lifetimes.    }

    \maketitle
%-------------------------------------------------------------------

\section{Introduction}

Over the past two decades, vortical motions have been extensively detected in the solar atmosphere due to the high resolution observations acquired by novel solar telescopes both from the ground and from space. This type of motions had already been predicted by theoretical calculations \citep{Norlund85} and were later confirmed by numerical simulations \citep{Moll12,Shel11,Kitia12,Shel13,Kato17,Yadav21}. In the photosphere, vortical motions appear to occur on the downdrafts of intergranular lanes. Since their first detection \citep{Brandt88}, several such motions have been revealed in observations. At chromospheric heights, where the magnetic field is potentially twisted by associated photospheric vortices, they appear as dark circular and spiral shaped patches called ``chromospheric swirls" \citep{Wede09}. The interest on vortex flows and chromospheric swirls has increased in recent years as they are considered potential candidates for contributing to the heating of the upper solar atmosphere. Consequently, in order to investigate the correlation between photospheric vortex flows and chromospheric swirls, as well as their role in the energy transfer between the corresponding atmospheric layers, it is necessary to have access to reliable values for their abundance and properties based on a sufficiently large statistical sample.

Visual inspection methods were initially used to study the properties of individual vortices and chromospheric swirls \citep{Bonet08,Balmaceda10,Wede12}.
In the latter study, there were 14 chromospheric swirls identified in observations obtained by the CRisp Imaging Spectro-Polarimeter (CRISP; \citealp{Schar08}) mounted on the Swedish 1-m Solar Telescope (SST; \citealp{Schar03}) in the \ion{Ca}{II}\,8542\,{\AA} spectral line. Larger statistical samples were obtained and examined after the development of automated detection methods. These methods have mainly been applied to photospheric observations, and to simulations as well. These use local correlation tracking (LCT; \citealp{Novemb88}) or Fourier local correlation tracking (FLCT; \citealp{Fisher08}) to derive the horizontal velocity fields in an attempt to automatically identify vortices in theses fields through vorticity or methodologies based on the velocity gradient tensor. \citet{Kato17} combined velocity fields with two different vorticity criteria (vorticity strength and enhanced vorticity) to automatically identify chromospheric swirls in three dimensional numerical simulations of the solar atmosphere and derive some of their statistical properties. In addition, FLCT and an automated vortex identification method based on $\Gamma$-functions was used by \citet{Giagkio17} on photospheric CRISP/SST \ion{Fe}{I} continuum observations to determine the centers and boundaries of vortices and some statistical properties. The same approach was used in the work of \citet{Liu19}, applied to time series of SOT/Hinode \ion{Ca}{II}\,H observations taken with a wide bandwidth filter to derive similar statistical information on swirls. The LCT method was also implemented in detection methods that take into account the variance of the vortex boundary in a time-varying velocity field. This approach was used in the work of \citet{Silva18}, where the authors made use of the Lagrangian-averaged vorticity deviation (LAVD; \citealp{Haller16}) to objectively identify vortex boundaries. An LAVD-based approach was also used to detect kinematic vortices in a 3D magnetohydrodynamic (MHD) dynamo simulation by \citet{Rempel17} who proposed a new method for the detection of magnetic vortices and the determination of their boundaries based on the integrated averaged current deviation (IACD) integral. This method was later applied by \citet{Rempel19} on a series of 2D and 3D simulations.

All the aforementioned methods have produced estimations for a number of significant physical properties of vortex flows and chromospheric swirls over recent years. \citet{Bonet08} provided the first estimations of such properties of small-scale photospheric vortex flows and obtained values of 1.8$\times 10^{-3}$\,vortexes Mm$^{-2}$\,min$^{-1}$ for the space-time density and a mean lifetime of 5.1\,$\pm$\,2.1\,min. Thanks to a visual inspection of magnetograms, \citet{Bonet10} provided updated estimations of a mean duration of $\sim$7.9\,minutes and a space-time density of 3.1$\times 10^{-3}$\,vortexes Mm$^{-2}$\,min$^{-1}$, along with an estimated mean radius of 0.5\,Mm. Similar values of mean radii were derived by \citet{Balmaceda10} and \citet{Vargas11}, 0.9\,Mm and 0.25\,Mm, respectively, while the mean lifetimes were found larger than previous estimations, namely, $\sim$10 - 20\,min in both works. The latter study provided also an estimation of space-time density of $\sim$1.4$\times 10^{-3}$\,vortexes Mm$^{-2}$\,min$^{-1}$. Individual cases of large-scale long-lived vortices in photospheric observations were studied by \citet{Brandt88}, \citet{Attie09}, and \citet{Reque18}. These authors reported radii of 2.5\,Mm, 7.5\,Mm and 2.5\,Mm, respectively, along with lifetimes greater than 1\,h. By applying their automated detection technique in observations of the photosphere \citet{Liu19} identified an average of $\sim$22 swirls per frame within a $\sim$800\,Mm$^{2}$ field-of-view (FoV) with an average radius of $\sim$0.29\,Mm. In the first report of a chromospheric swirl, \citet{Wede09} derived a radius of 0.75\,Mm. \citet{Wede12} by studying a larger sample of 14 swirls observed in the chromosphere derived a mean lifetime of 12.7\,min and a space density of 2$\times 10^{-3}$Mm$^{-2}$. \citet{Park16} revealed, for the first time in H$\alpha$ observations, two swirling events with radii of 0.3 and 0.7\,Mm and lifetimes of 1 and 2\,min, respectively. \citet{Tzio18} studied a large-scale persistent chromospheric swirl in a quiet region and derived a radius of $\sim$2.2\,Mm and a lifetime of at least 1.7\,h. Radii of $\sim$1\,Mm and lifetimes of $\sim$10\,min were also derived by a multiwavelength investigation of 13 swirls in the study by \citet{Shetye19}. The automated detection method of \citet{Kato17} (mentioned above) resulted to an estimated occurrence rate of roughly 1 vortex\,Mm$^{-2}$\,min$^{-1}$, which implies the existence of a chromospheric swirl for every granular cell. Additionally, by isolating the large-scale vortex flows with diameters larger than 1\,Mm, they derived occurrence rates of 7.1$\times 10^{-2}$\,vortices\,Mm$^{-2}$\,min$^{-1}$ and 8.8$\times 10^{-3}$\,vortices\,Mm$^{-2}$\,min$^{-1}$ from the two different vorticity-based criteria. We refer to Table \ref{tab:1} for a summary of parameters from previous studies (the results of the present study are also included for comparison).
In this paper, we attempt to detect the existing swirls in the entire FoV in a time series of high-resolution, high-cadence H$\alpha$ observations and extract statistical information for a number of critical physical parameters of these structures. The detection was performed via an automated detection method of chromospheric swirls presented by \citet{Daka21} (hereafter Paper I) which exploits the morphological characteristics of the observational signature of swirling motions in the layer of the chromosphere. The rest of the paper is organized as follows: in Section \ref{sec:obs}, the set of observations used in the present analysis is described. In Section \ref{sec:method}, a brief description of the automated detection method is given. In Section \ref{sec:stats}, we discuss its application for the new dataset. In Section \ref{sec:results}, we present estimations of various parameters of the detected individual swirls, as well as those derived from a statistical analysis. Finally, we present our discussion and conclusions in Sections \ref{sec:disc} and \ref{sec:concl}.

\section{Observations and data reduction}\label{sec:obs}

The time series of H$\alpha$ observations used in the present work is part of a multi-wavelength, multi-line datasets obtained during a coordinated campaign. It was taken on August 13, 2019 with the SST/CRISP. The aim of the observing campaign was the observation of swirling phenomena in the quiet Sun at different atmospheric heights and the target a quiet region at the solar disk center. CRISP sampled the H$\alpha$ spectral line with a narrowband filter of 0.061\,{\AA} at 13 equally spaced wavelengths around the line center at 6563\,{\AA} in the spectral window [$-$1.2\,{\AA}, 1.2\,{\AA}] with 0.2\,{\AA} steps, as well as the H$\alpha$ line using a wideband filter of 4.9\,{\AA}. The CRISP instrument covers a 57$\arcsec \times$ 58$\arcsec$ FoV with an image pixel scale of 0.0592$\arcsec$.
\begin{figure*}[htb!]%%%%%%%%%%%%%%%%%% FIGURE 1 - dataset images
\center
{\includegraphics[width=0.259\textwidth]{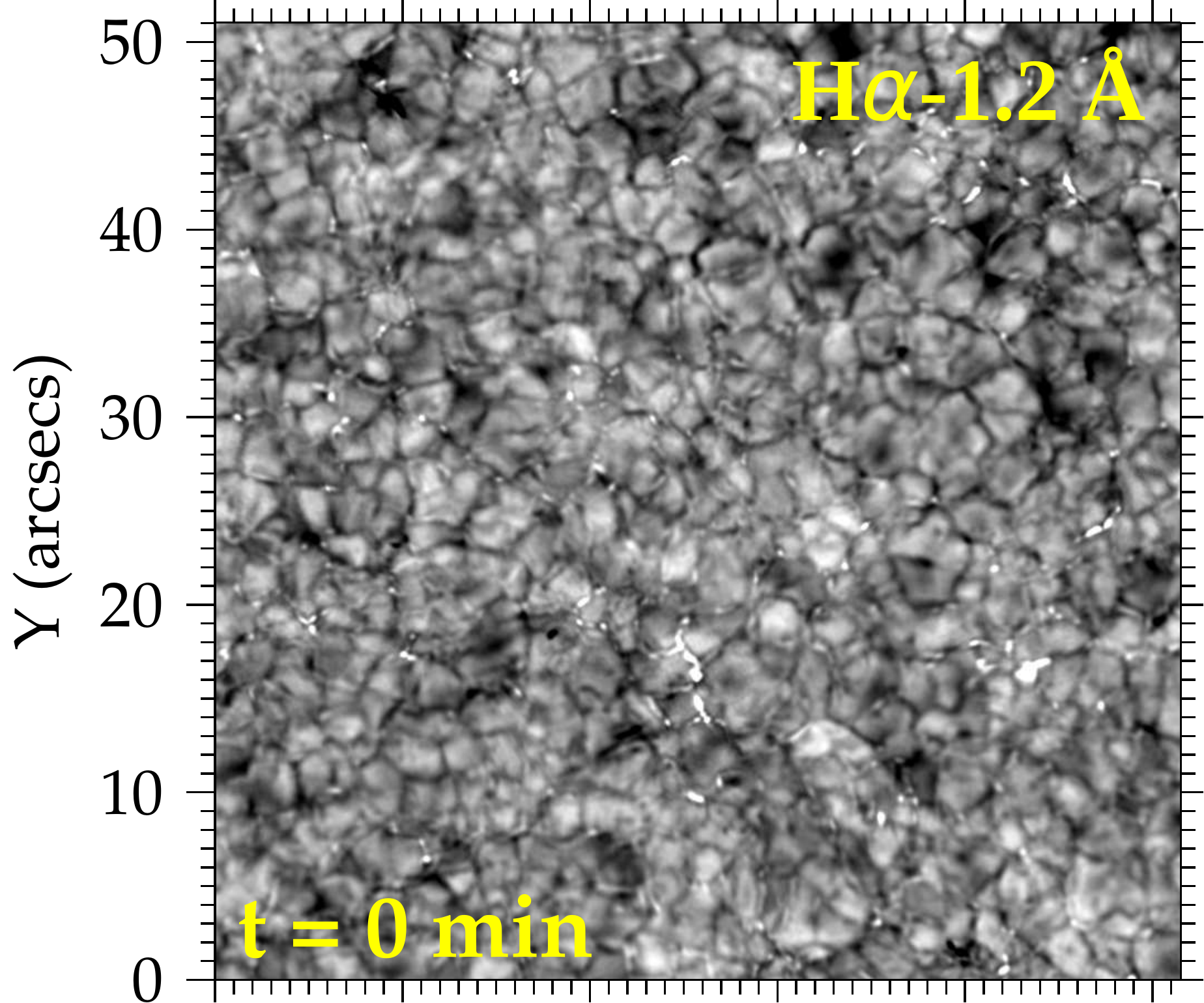}}
{\includegraphics[width=0.22\textwidth,clip=,bb= 10 0 461 443]{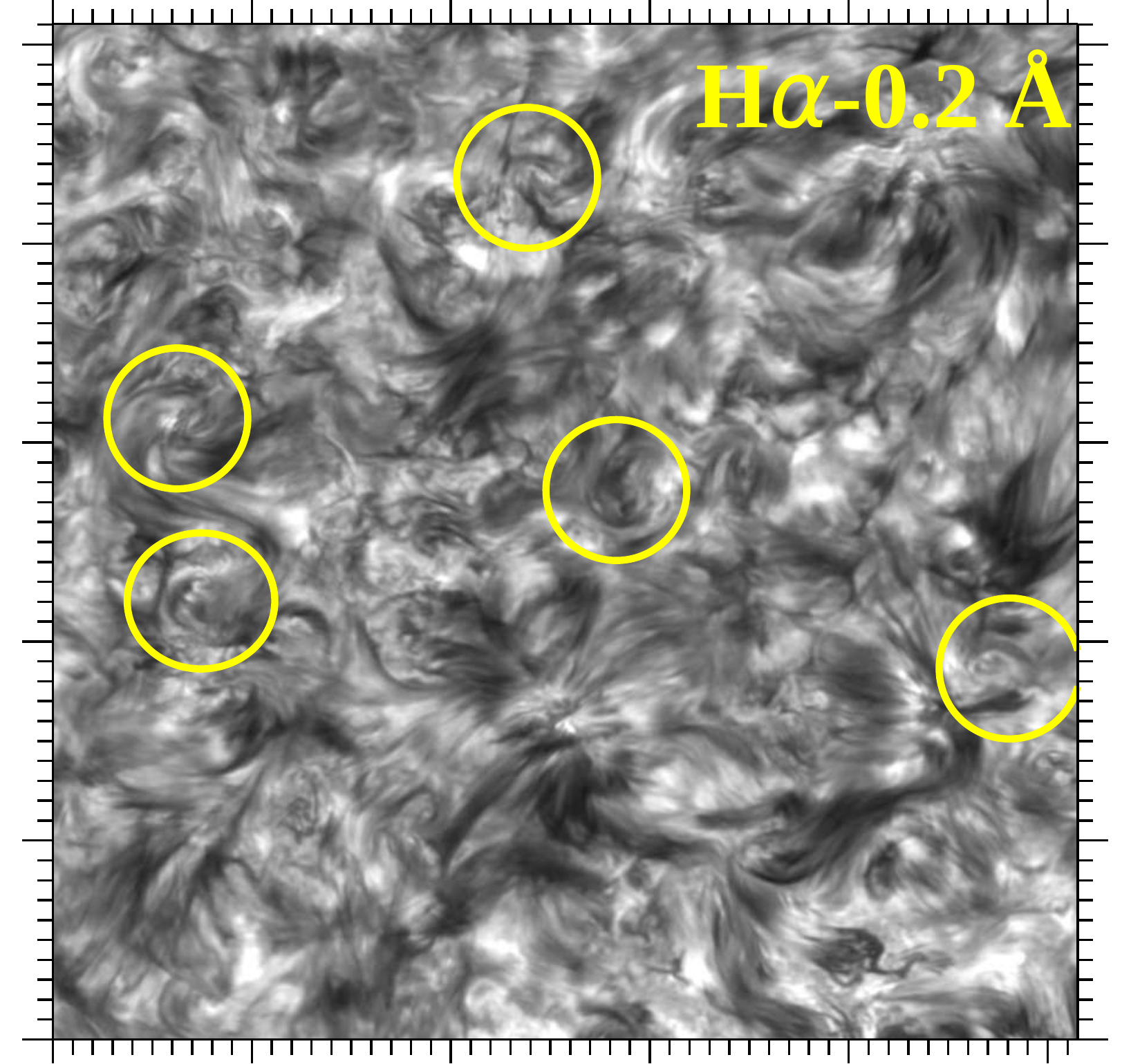}}
{\includegraphics[width=0.22\textwidth,clip=,bb= 10 0 461 443]{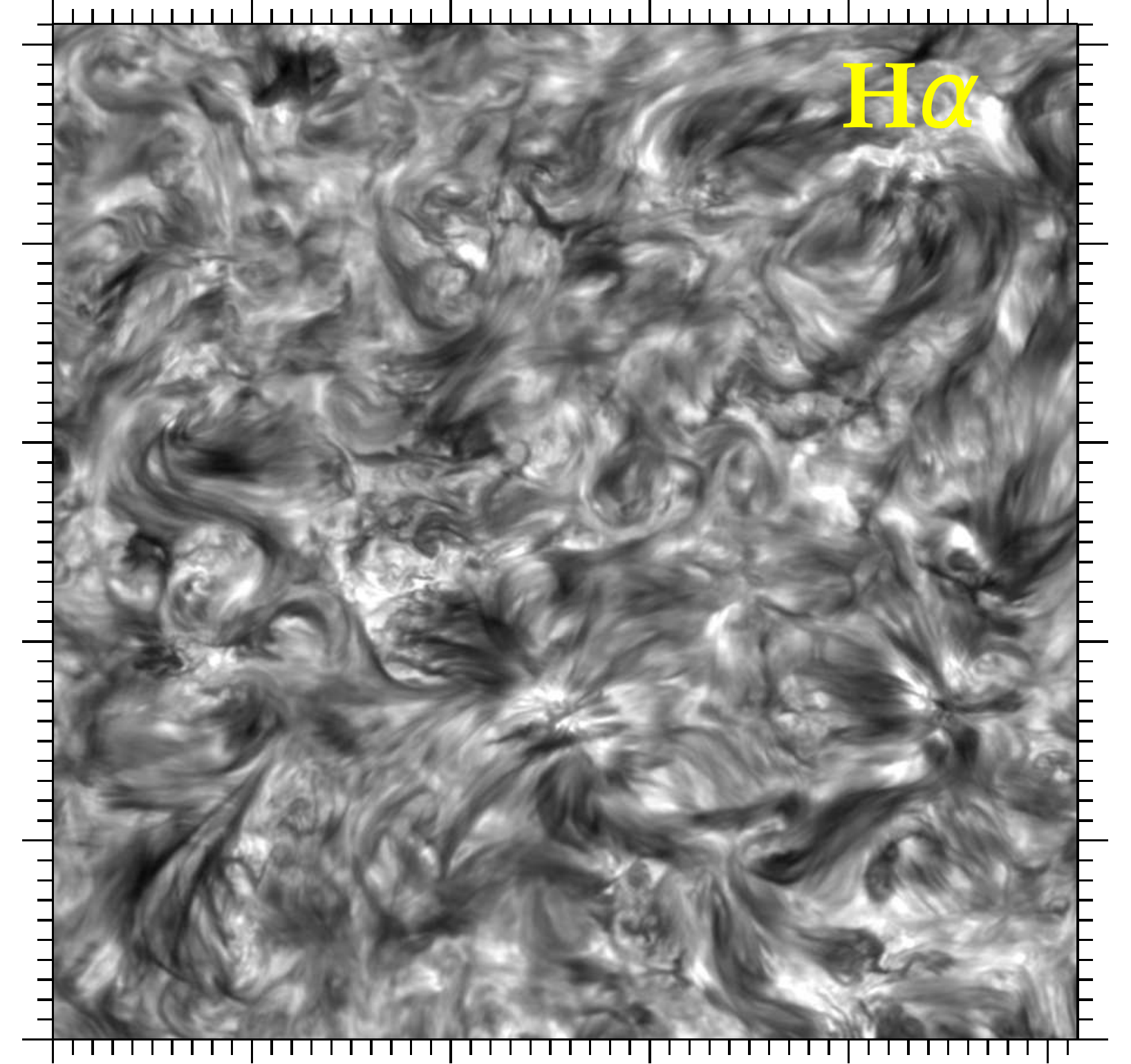}}
{\includegraphics[width=0.22\textwidth,clip=,bb= 10 0 461 443]{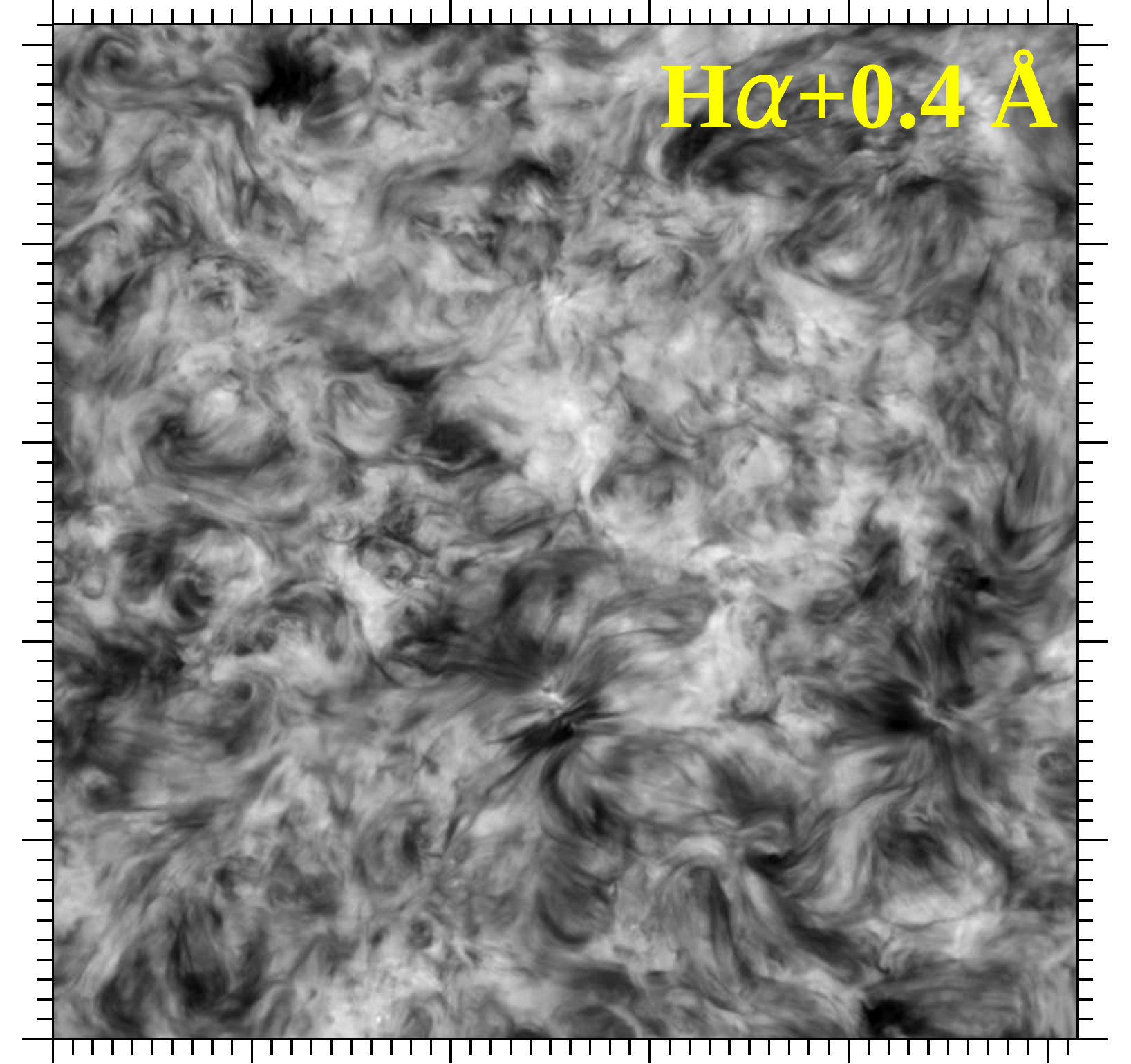}}
{\includegraphics[width=0.259\textwidth]{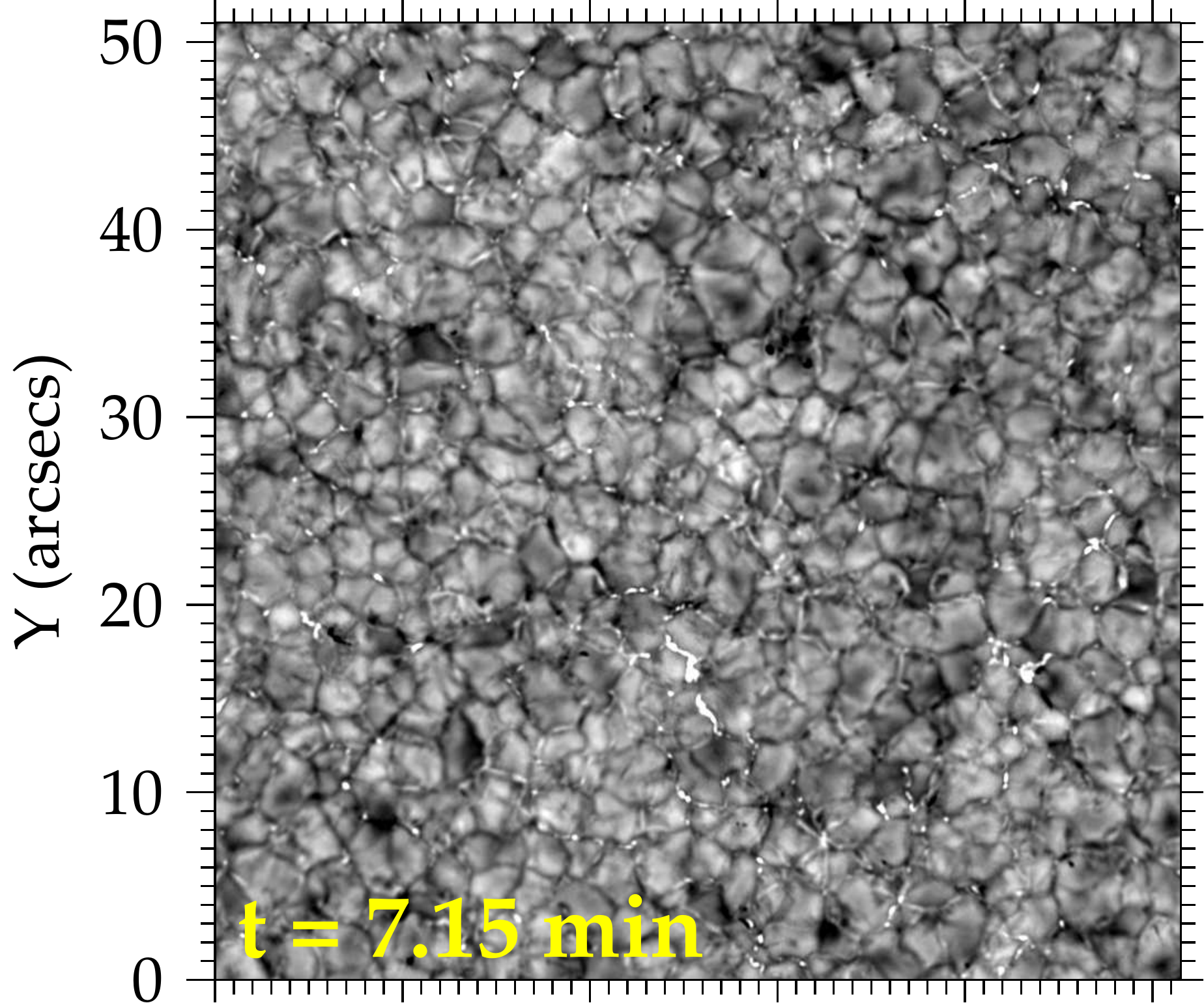}}
{\includegraphics[width=0.22\textwidth,clip=,bb= 10 0 461 443]{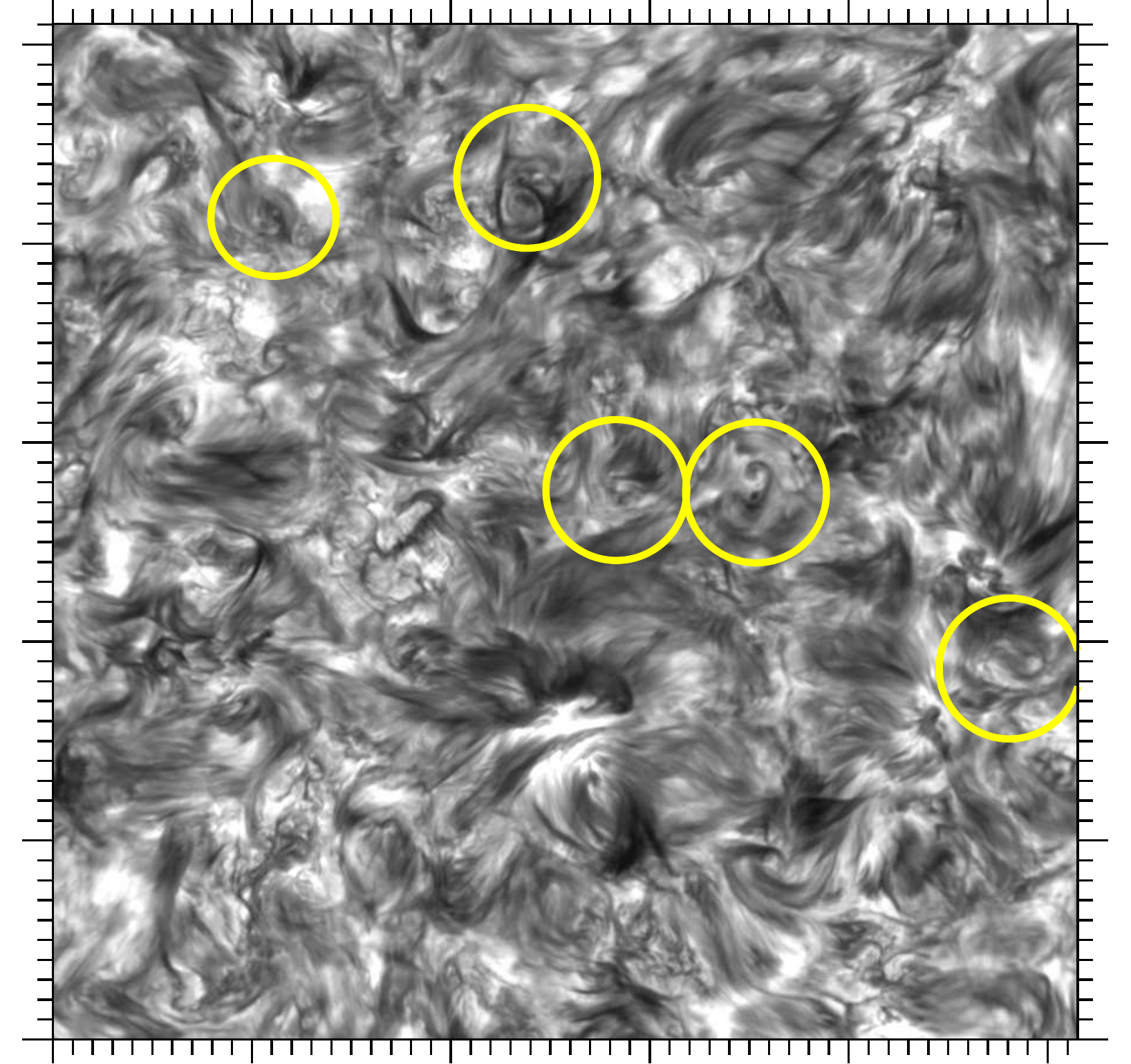}}
{\includegraphics[width=0.22\textwidth,clip=,bb= 10 0 461 443]{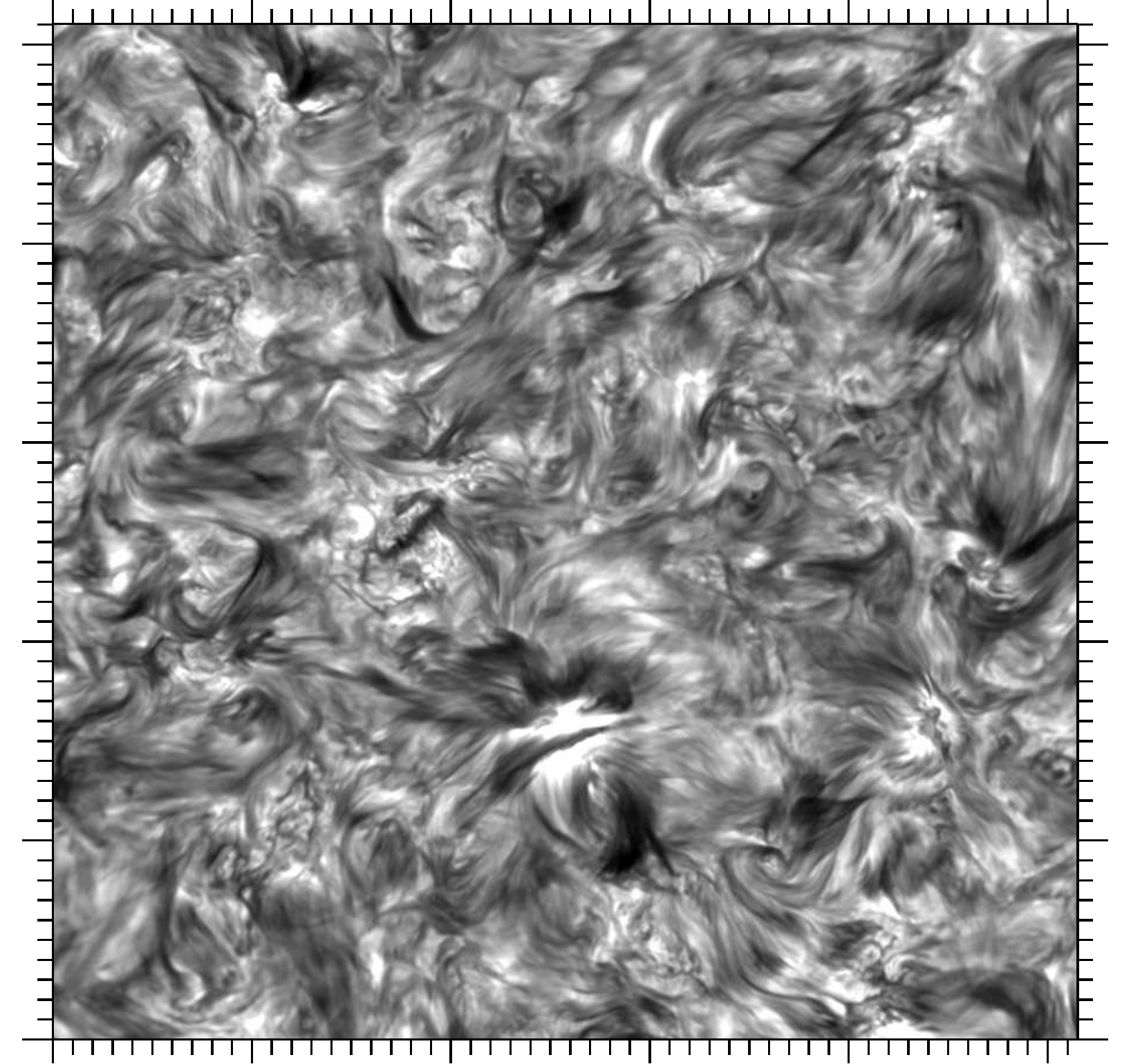}}
{\includegraphics[width=0.22\textwidth,clip=,bb= 10 0 461 443]{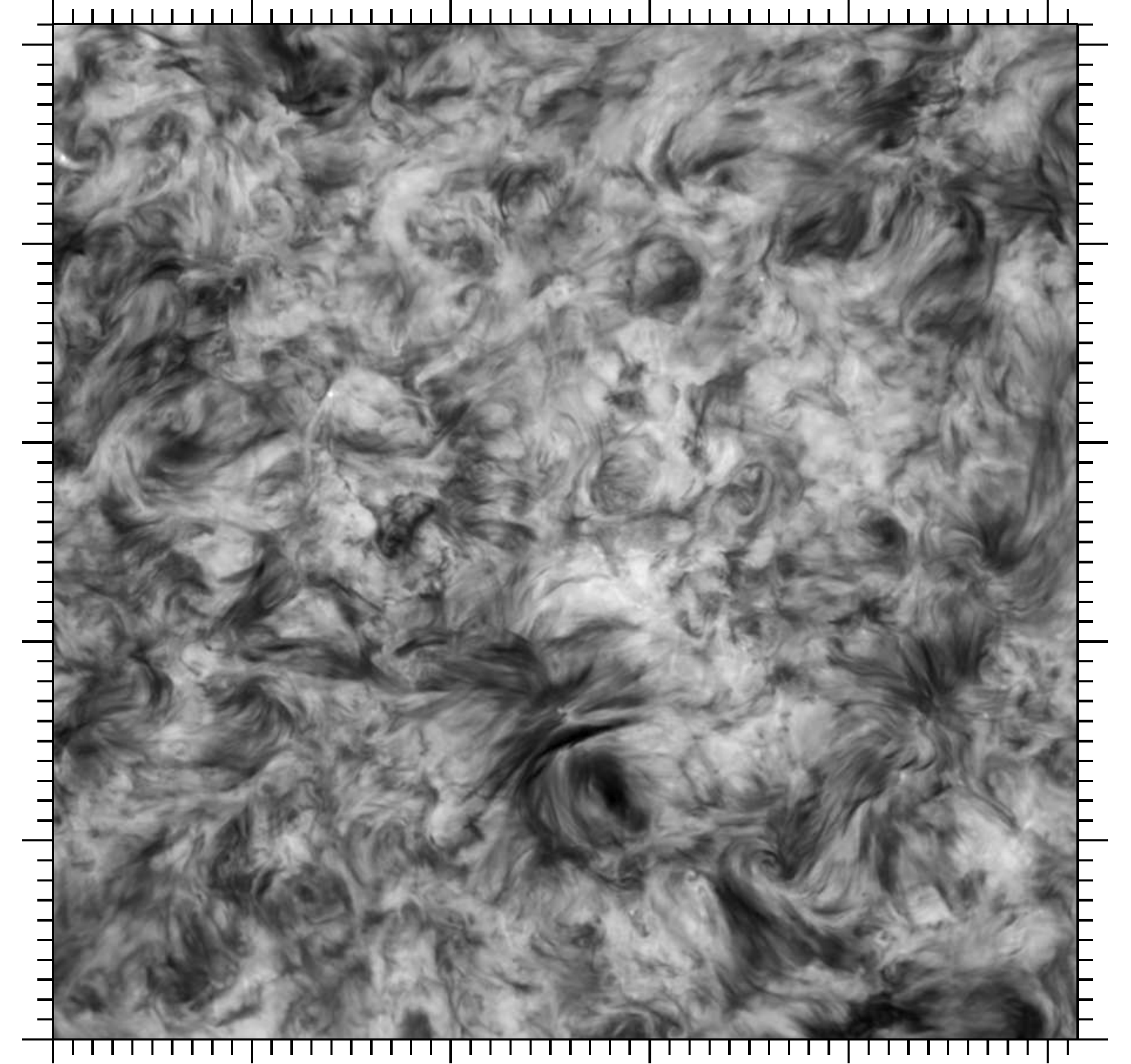}}
{\includegraphics[width=0.259\textwidth]{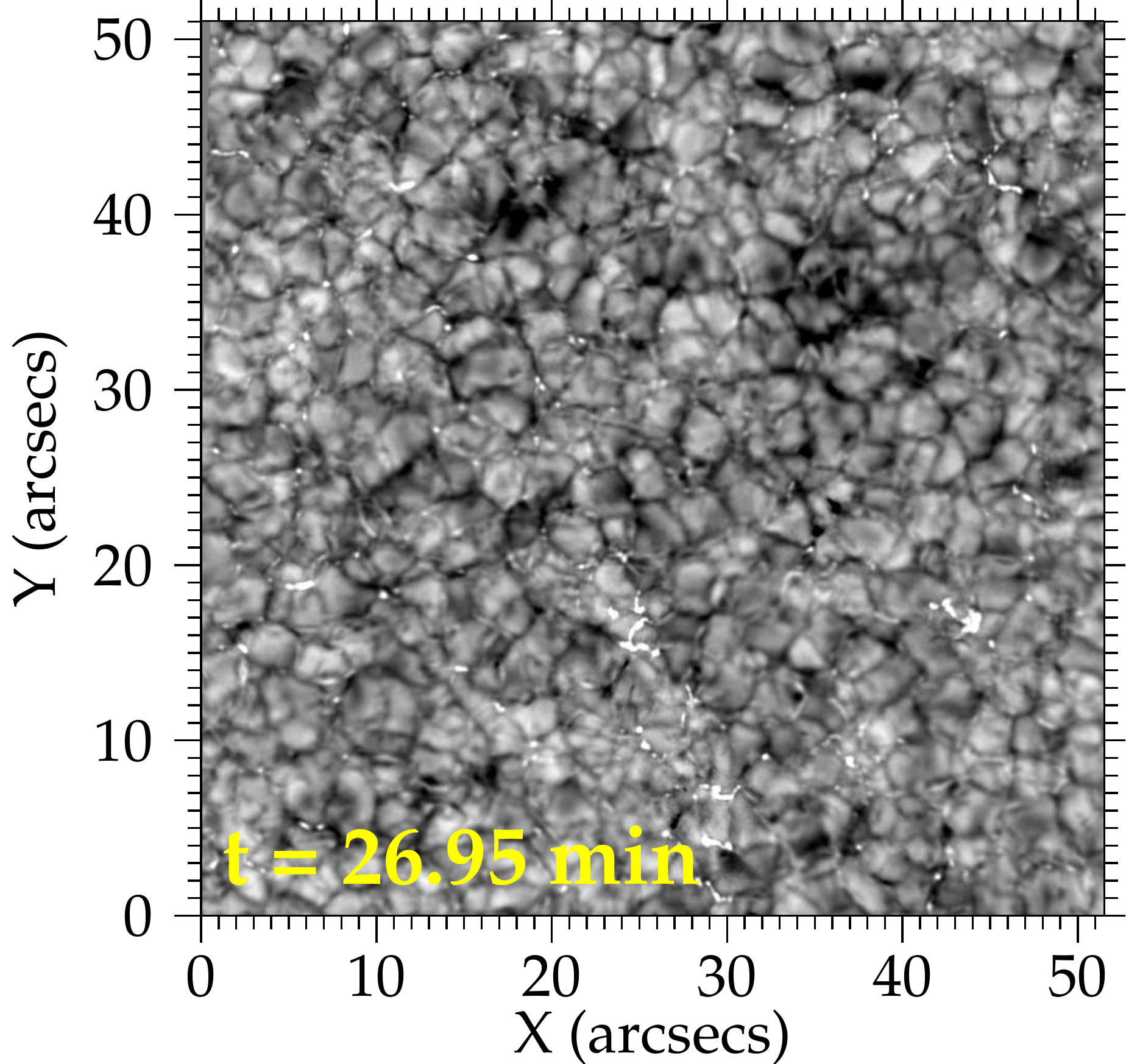}}
{\includegraphics[width=0.22\textwidth,clip=,bb= 10 0 461 503]{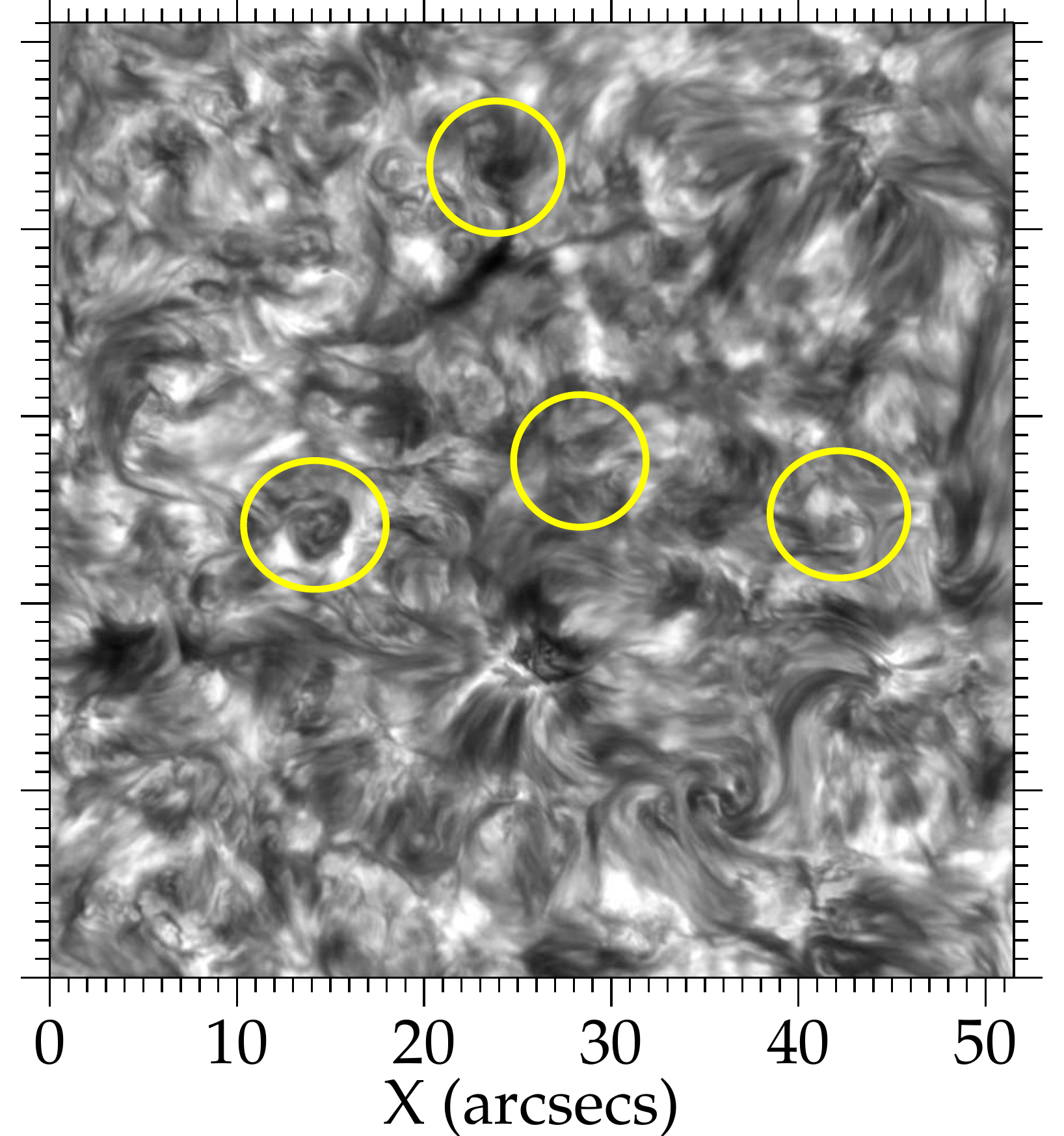}}
{\includegraphics[width=0.22\textwidth,clip=,bb= 10 0 461 503]{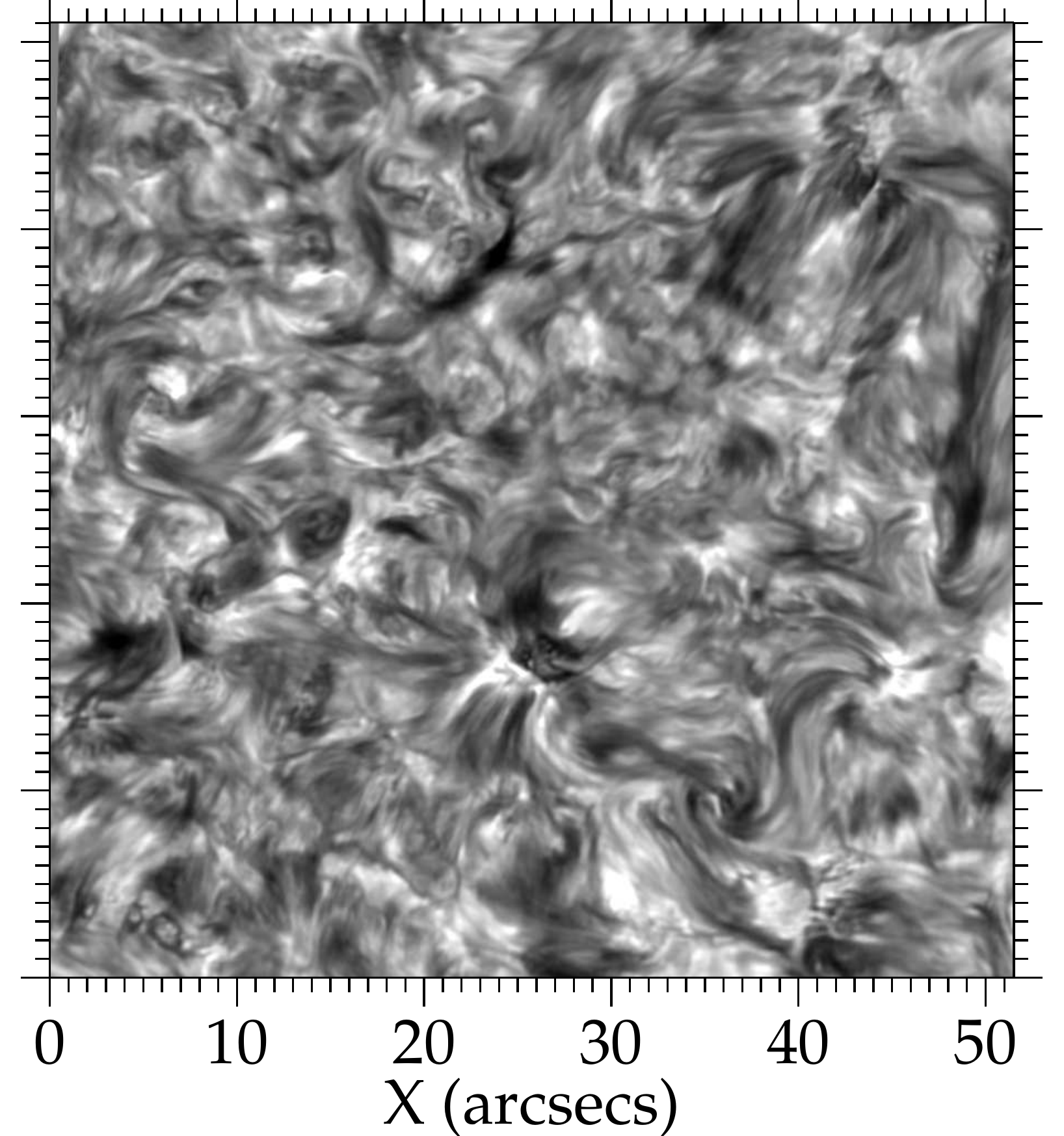}}
{\includegraphics[width=0.22\textwidth,clip=,bb= 10 0 461 503]{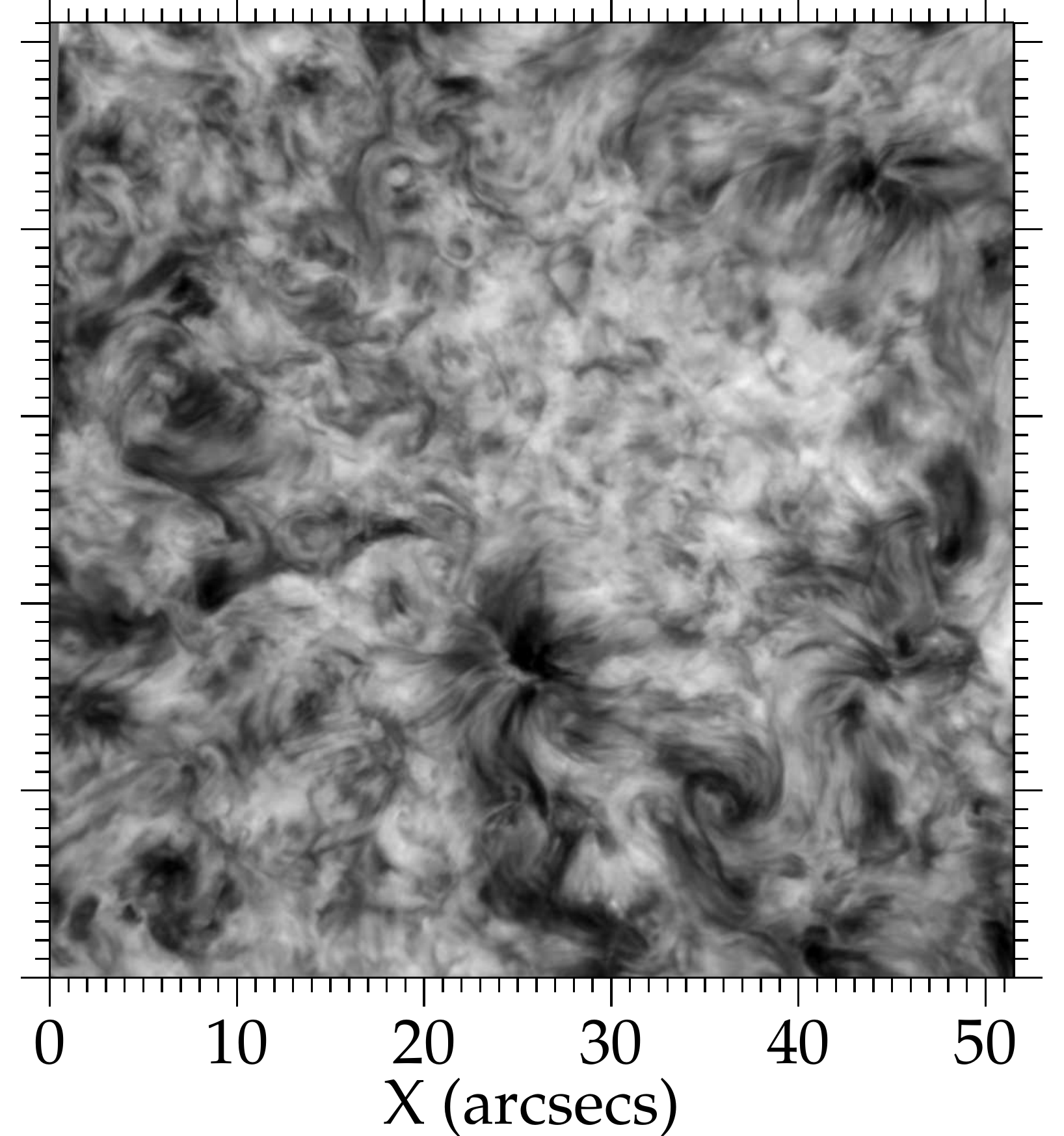}}
\caption{Overview of the SST/CRISP H$\alpha$ observations analyzed in this study. Time $t=0$ corresponds to 10:20 UT (start of the observations). Each row shows snapshots of intensity maps taken at H$\alpha$-1.2\,{\AA}, H$\alpha$-0.2\,{\AA}, H$\alpha$ line center and H$\alpha$+0.6\,{\AA}, respectively, while each column corresponds to same-wavelength snapshots at different times of the time sequence. The yellow circles in the second column denote several conspicuous cases of swirls, with the most persistent ones visible in more than one snapshot. An animation of the entire H$\alpha$-0.2\,{\AA} time sequence clearly displaying the large number of swirls in the online journal.  }
\label{fig:observ}
\end{figure*} The H$\alpha$ time sequence runs from 10:20:37 UT to 10:54:24 UT with an average cadence of $\sim$6.3\,s and 320 images in each wavelength out of which 16 were excluded during the initial pipeline because they were highly blurred. After $\sim$25\,min from the beginning of the observations, there is a slight displacement of the pointing, resulting in a rectangular common FoV of the two darasets equal to 51.3$\arcsec \times$ 53.6$\arcsec$. Simultaneous datasets were also acquired in the \ion{Ca}{II}\,IR spectral line with the same instrument, in the \ion{Ca}{II} H and K spectral lines with the CHROMospheric Imaging Spectrometer (CHROMIS; \citealp{Schar17}), as well as in various UV and EUV lines together with SJ images with the IRIS space mission. In the present study, however, only the H$\alpha$ dataset is used. In a follow-up work, a multi-wavelength analysis will be performed.

The raw data were processed using the CRISPRED reduction pipeline (\citealp{delaCruzRodriguez15}). The pipeline includes image restoration via Multi-Object Multi-Frame Blind Deconvolution (MOMFBD; \citealp{vanNoort05}) to remove the atmospheric distortion. In this process, the reconstructed wideband images serve to set the precise alignment of the narrowband images. The very good seeing conditions, for most of the observation time, the SST adaptive optics system, and the restoration pipeline provide a high quality observational dataset. The FoV is covered with a large number of conspicuous, persistent swirling phenomena. These are more clearly visible at H$\alpha$ line center and near H$\alpha$ line center wavelengths (Fig. \ref{fig:observ}). In the provided online movie, the presence of a large number of swirls in the observed FoV is ubiquitous.

\section{Description of the automated swirl detection method and parametrization of the algorithm}\label{sec:method}

In order to carry out a meaningful statistical analysis and derive reliable values of the parameters of chromospheric swirls, a sufficiently large sample of observational data is essential. Fortunately, the presence of a large number of swirls in each frame and the relatively long duration of the current H${\alpha}$ observations provide a unique opportunity for such an analysis. Indeed, based on a visual inspection of the H$\alpha$ filtergrams, even at a first glimpse, many swirling areas can be identified on wavelengths close to the H$\alpha$ line center (Fig. \ref{fig:observ}). In order to automatically detect the existing swirls in the filtergrams and provide reliable values for their physical properties and their populations, a novel chromospheric swirl detection algorithm was developed (presented in Paper I). It detects swirling events in high-contrast images of the solar chromosphere by exploiting their morphological characteristics. The algorithm is applied to the H$\alpha$-0.2\,{\AA} images where swirls are better visible than in the H$\alpha$ line center (see also \citealp{Tzio18}) because the FoV is less covered by the dark chromospheric fibrillar structures and, in addition, it contains an upward velocity signature that is observed in swirls' arms. It consists of four standalone stages: (i) image pre-processing, (ii) tracing of highly curved structures, (iii) selection and combination of traced dark, curved segments, and (iv) a two-tier clustering. The latter involves clustering (a) in space providing the swirl candidates and, subsequently, (b) in time providing the detected, persistent swirls. A brief description of the method is given below, while  an extensive description is given in Paper I.

In the first stage, the difference of Gaussians (DoG) edge enhancement filter is applied to each H${\alpha}$-0.2\,{\AA} image of the dataset. As a result, a large amount of noise is excluded and the values of the intensities of each filtered image form a Gaussian-like histogram that is fitted with a Gaussian probability density function. Then intensity values are scaled between $[\mu-4\sigma,\mu+4\sigma]$ to increase the contrast of the image, where $\mu$ is the mean value and $\sigma$ is the standard deviation of the best fitted function.
\begin{figure*}[htb!]%%%%%%%%%%%%%%%%%% FIGURE 2 - barplot
{\includegraphics[width=\textwidth]{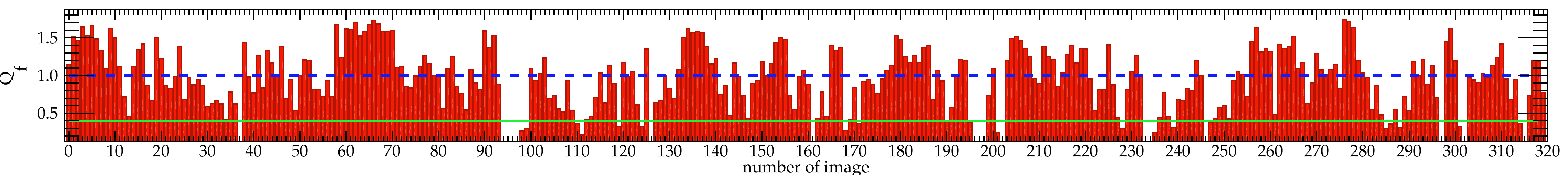}}
\caption{Barplot of the $Q_{f}$ values of the images of the H$\alpha$-0.2\,{\AA} dataseries. The horizontal dashed blue line corresponds to images with a ratio of $Q_{f}=1$, namely, images with the mean quality. Images with $Q_{f}$ less than 0.4 were considered as highly blurred and were removed. This threshold is marked by the horizontal light green line. The gaps in the plot correspond to a total number of 21 removed images from the initial pipeline (16 images) and the application of the above threshold (five more images).  }
\label{fig:fwhm}
\end{figure*}
In order to asses the quality of each image, the ratio, $Q_{f}$, of the full-width-at-half-maximum (FWHM) of the Gaussian distribution of its intensities to the mean FWHM of the entire time sequence is calculated. The first stage concludes with an adaptive (local) intensity thresholding at each image in order to highlight dark areas for the tracing stage. The obtained masks contain dark structures, namely, areas with negative flux surrounded by zero-valued areas as the upper threshold.

The second stage, namely, the tracing component of the algorithm, starts from the minimum flux point of the entire image, which, as a consequence of the previous stage, is negative and part of a dark structure. The structure is then sequentially traced using a negative flux-based directivity criterion until a zero boundary value is reached. Subsequently, the traced structure is erased, the algorithm locates the next minimum flux point and repeats the process for the corresponding structure. During the tracing of each structure, a local curvature criterion is applied on each segment, which is based on the standard deviation of the angles of the segment points in order to retain only the highly curved structures.

In the third stage, a global curvature criterion and a minimum radius criterion are applied on the traced segments. The minimum radius, $r_{min}$, was set based on the assumption that no chromospheric swirls with a radius less than 200\,km, have been reported to date. Moreover, the tracing of unreliable structures close to the pixel scale of CRISP images which is 0.059$\arcsec$/pixel$\simeq$42.8\,km/pixel is prevented. The remaining segments are combined over a number of consecutive images, $T$, which is chosen depending on the cadence of the dataset as, $max[cadence\cdot T]\leq 20$\,s, $T\in 2\mathbb{N}+1$. In particular, for each i$^{th}$ image of the dataset we consider the $\sum_m N_{m}$ traced and retained segments of $T$ consecutive images, where $m=i-(T-1)/2,..i,..i+(T-1)/2$ and $N_{m}$ is the number of traced curved segments of the m$^{th}$ image.

The coordinates of these segments along with the centers of curvature are then passed to the fourth stage of the algorithm. An unsupervised machine learning technique is used in order to cluster together centers of curvature that form high point density areas. In particular, the density-based spatial clustering of applications with noise (DBSCAN; \citealp{Ester96}) method creates clusters of centers of curvature in each image, which are labeled as Swirl Candidates (SCs). The result depends on two key parameters: the radius $\epsilon$ of the local circle around a point and the $MinPts$ parameter, which indicates the minimum number of points that must reside within $\epsilon$ in order for the points to be considered as part of the same cluster. This stage consists of two levels of clustering. In the first level clustering the radius is set as $\epsilon=2r_{min}$, where $r_{min}$ is the minimum swirl radius constraint set earlier, while $MinPts$ is fixed and equal to the $T$ interval. In the second level clustering the same algorithm is used to cluster together the two dimensional projection of the median centers of the SCs from every moment in time, namely, from every image of the dataset, creating the final clusters of structures, which are characterized as ``swirls.'' The $\epsilon$ parameter remains the same at this level but $MinPts$ changes to $\lfloor t_{sc,min}/cadence \rfloor$ based on the assumption that an SC must preserve a center of curvature within $\epsilon$ for at least a total of $t_{sc,min}=1.5$\,min for (at least) a 5 min duration data series in order to be considered a genuine swirl.

\section{Methodology and analysis of the data}\label{sec:stats}

\subsection{Automated detection and tracking of swirls}\label{subsec:detection}

We applied the algorithm presented analytically in Paper I (and described in brief in the previous section) to the current H$\alpha$-0.2\,{\AA} observations, which have a duration of 33.7\,min. In order to be able to automatically reveal the low quality images, the FWHM-based quality determination process of the first stage of the algorithm was used. The $Q_{f}$ criterion was calculated for every image and images with a $Q_{f}<0.4$ were considered to be of low quality and were thus removed from the sample (Fig. \ref{fig:fwhm}). As a result, five more low-quality images were removed during this process. Visual inspection also allowed us to verify the poor quality of the removed images. All gaps in the data series were replaced by zero valued images of the same dimensions. The parametrization, up to the first level of clustering, was identical to the one used in Paper I. However, the minimum radius parameter of the second level clustering was modified to $\epsilon\simeq r_{min}\simeq 340$\,km due to the increase of detected centers that are provided to the clustering stage, which is, in turn, an effect brought on by the increased quality and length of the present dataset compared to the previous one.

\begin{figure*}[htb!]%%%%%%%%%%%%%%%%%% FIGURE 3 -3D plot
{\includegraphics[width=0.95\textwidth]{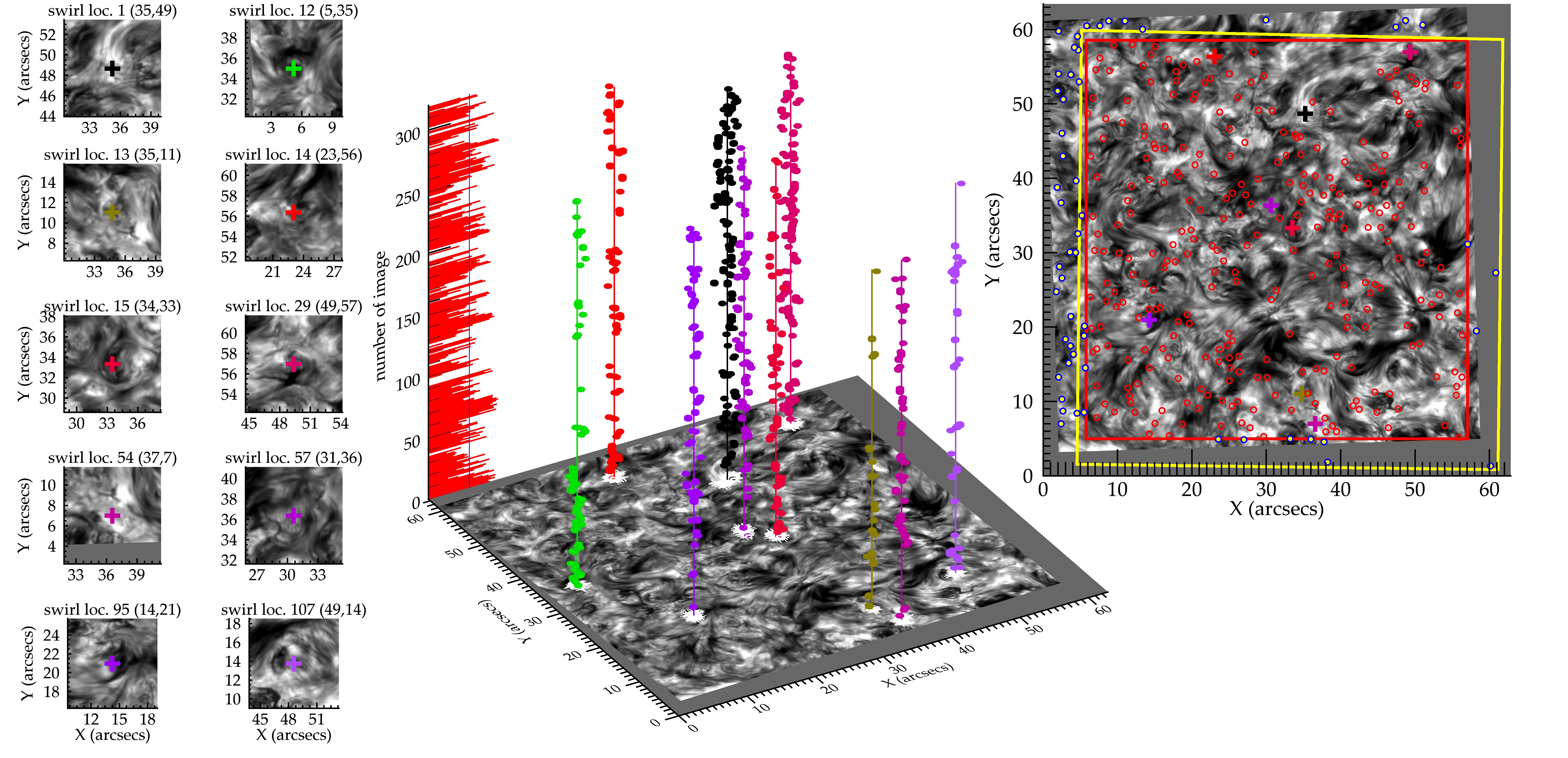}}
\caption{Spatial and temporal evolution of the detected swirl centers. The panels at the left part of the figure correspond to snapshots of 10 out of 350 randomly selected swirl locations detected by the algorithm, as they appear in the first image of the H$\alpha$-0.2\,{\AA} time series. The colored crosses represent the overall center of each swirl location. The 3D plot in the middle depicts the temporal evolution of the centers of the co-located swirling events, as derived by the second level clustering which reside within these 10 locations prior to the application of the adopted assumptions (see text). The FWHM barplot is overplotted on the z-axis for cross-reference of low quality or missing images. The right panel shows the entire FoV of the first image of the H$\alpha$-0.2\,{\AA} dataseries, with all the detected 350 swirling locations, along with the borders of the FOV after the displacement (yellow rectangle) and the common FoV (red rectangle). The 10 swirl location centers are overplotted as colored crosses along with the remaining swirl locations both within (red circles) and outside of (blue circles) the common FoV.  }
\label{fig:results}
\end{figure*}

The application of the algorithm to the dataset resulted in the detection of 6161 swirling events (irrespective of their duration) at 350 locations, suggesting the presence of several co-located swirling events. The spatial and temporal evolution such of swirling events in ten randomly selected locations is presented in Fig. \ref{fig:results}. The middle panel of the figure is constructed by matching each mean center of curvature of the final second level clustering to the corresponding image of the dataset. As it is seen in Fig. \ref{fig:results} (middle and right panels), the locations of the detected swirling events are not distributed homogeneously in the FoV and are less frequent in areas where linear fibrillar structures reside. Their relation to photospheric bright points, as proposed by \citet{Wede09} and \citet{Wede12}, requires further investigation that is beyond the scope of this paper.
\begin{figure}[htb!]%--------------- FIGURE 4 - filling interval
%\resizebox{\hsize}{!}
{\includegraphics[width=0.45\textwidth]{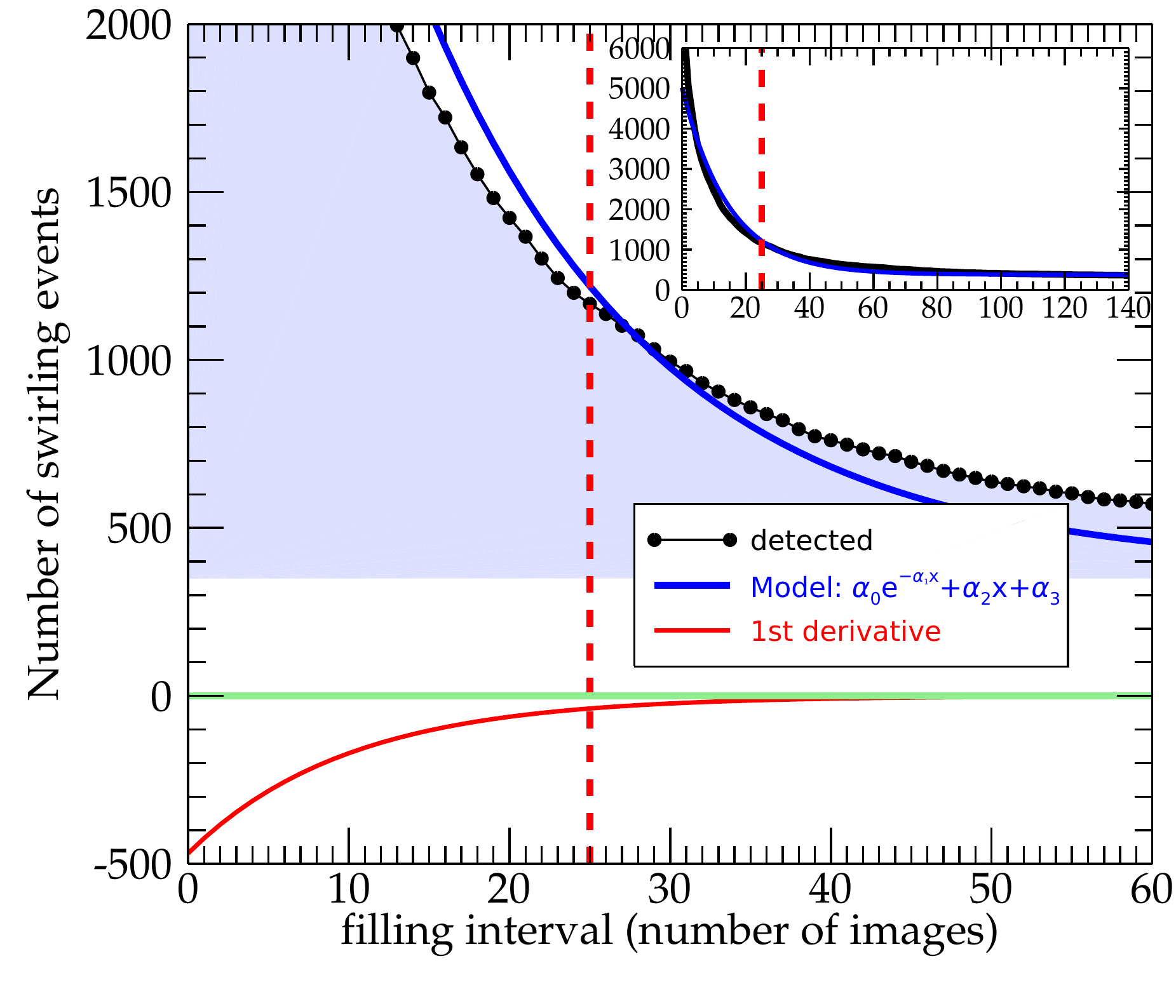}}
\caption{Number of detected swirling events after the application of the second level clustering vs. the number of gap-filling frames. In the inset figure, we show that the total number of detected swirling events decreases abruptly with the number of frames and finally converges to a constant value. At the zoom-in of the inset figure the fitting curve (blue line) of the detected one is plotted (black line) as well as its first derivative (red line). The dashed line shows the estimated number of the gap-filling frames. The lower border of the shaded area shows that the number of detected swirling events naturally converges to a number of 350 swirls that is equal to the derived locations when all gaps are filled in thanks to the respective large filling intervals.   }
\label{fig:model}
\end{figure}

It must be noted that apart from swirling events that can be followed between successive frames without disruption, there are, most of the time, also short or even long time gaps between the co-located swirling events, which are evident in their temporal evolution (see the middle panel of Fig. \ref{fig:results}). The observed time gaps may be due to: a) the formation mechanism, namely, either the swirling events are independent to each other or are recurrent and tend to reappear at the same place due to some as-yet-unknown physical process; b) changes of their internal physical properties with time; c) the dynamic nature of the chromosphere because of which even the most circular rotating structures at times become more elongated and therefore fail to pass the curvature criterion of the algorithm and their subsequent detection as swirling events; d) missing frames from the initial data series or frames removed during our processing (see Section \ref{sec:obs} and Fig. \ref{fig:fwhm}). Hence, an important question arises in the case of time gaps and its answer is vital to the statistical analysis of the various parameters; namely, it considers whether the swirling events appearing at the same location can be considered as the same swirl despite the time gaps or whether they have to be considered as individual swirls.
\begin{figure*}[htb!]%--------------- FIGURE 5 - lifespans
%\resizebox{\hsize}{!}
%{\includegraphics[width=0.95\textwidth]{lifespans_25merg_no13.pdf}}
{\includegraphics[width=\textwidth]{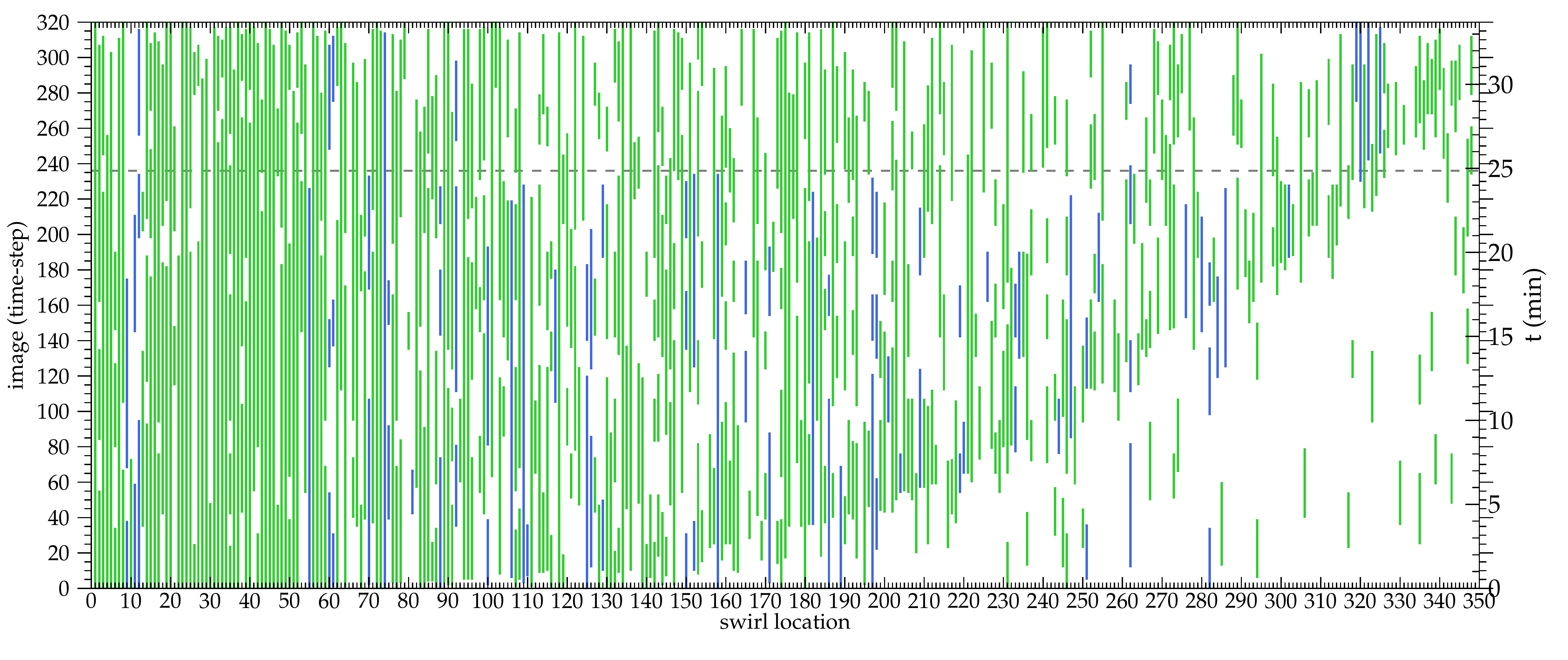}}
\caption{Temporal evolution of the 577 individual swirls in 350 swirl locations. The horizontal dashed line represents the time of the slight displacement of the FoV (see Fig. \ref{fig:results}). The green lines represent the temporal evolution of the 487 out of 577 swirls that lie within the selected common FoV. The blue lines represent the temporal evolution of swirls that cannot be followed through the entire time sequence due to the displacement of the FoV.    }
\label{fig:lifespans}
\end{figure*}

\subsection{Preparation of the sample for the statistical analysis}\label{subsec:preparation}

In order to overcome the drawback of time gaps between frames (for the reasons stated in Sect. \ref{subsec:detection}) on the estimation of the various parameters and to derive a well-defined statistical sample of features with common characteristics, a gap-filling procedure for co-located swirling events was considered. The number of frames for gap-filling should be large enough in order to eliminate time gaps due to selection criteria imposed by the algorithm or to missing frames; but it should also be small enough to ensure that co-located swirling events are not considered as the same swirl when a significant time gap exists between them. As Fig. \ref{fig:model} demonstrates, this number can be estimated by assuming that the number of discrete detected swirling events begins to converge toward a constant value above a certain number of gap-filling frames. According to this figure, a reliable value for the number of gap-filling frames to be used in the current observations is estimated from the first derivative of the analytical fit to the curve of detected swirling events. This value is equal to $\sim$25 frames (or given the cadence of the observations to 2.63\,min). This estimated value is deemed reliable and roughly equal to twice the minimum value of 1.5\,min for a structure to be considered as a swirl, which is one of the selection criteria of the detection algorithm (see Section \ref{sec:method}).
Therefore, from this point forth, our analysis is constructed upon three basic assumptions: (a) time gaps in the temporal evolution of the swirling events' centers less or equal to 25 frames are filled in and the swirling events are considered as the same swirl; (b) swirling events within the same location whose temporal evolution has a gap larger than 25 frames are considered as distinct swirls; (c) all swirling events that are detected in less or equal to 14 frames (i.e., for  $\la$ 1.5\,min) are removed from the sample in accordance to the selection criterion of the detection algorithm (see Sec. \ref{sec:method}).

Following these assumptions, the total number of 6161 detected swirling events is decreased to 588 individual swirling events which are distributed along the 350 swirl locations. It should be noted that in the H$\alpha$-0.2\,{\AA} FoV, apart from the swirling events, there is also a number of dark curved mottles that are apparent, with some of them included in this detected sample. To exclude them from the statistical sample, the standard deviation of the angles of the segments constituting the curved structures is calculated. The angle of each segment is defined as the angle between the distance of the central point of the segment with coordinates $(x_{j},y_{j})$, from the swirl center $(X_{s},Y_{s})$ and the horizontal. A swirling event with moderate or negligible rotation is expected to have a small standard deviation of angles and would most likely to be associated with a curved fibrillar structure. For 11 of the detected curved swirling events (i.e., $\sim$2\%), this value is less than 50$^{\circ}$; they were, thus, removed from the statistical sample. The temporal evolution of the remaining 577 swirling events, hereafter swirls, as well as their distribution throughout the 350 swirl locations is shown in Fig. \ref{fig:lifespans}.

\section{Results}\label{sec:results}

Following the strict delimitation of our statistical sample of swirls observed in H$\alpha,$ we can now derive meaningful statistical parameters, such as occurrence rate, lifetimes, radii, etc.

\subsection{Radii of swirls}\label{subsec:radii}

\begin{figure}[htb!]%--------------- FIGURE 7 - radii histogram
\resizebox{\hsize}{!}
{\includegraphics[width=0.33\textwidth]{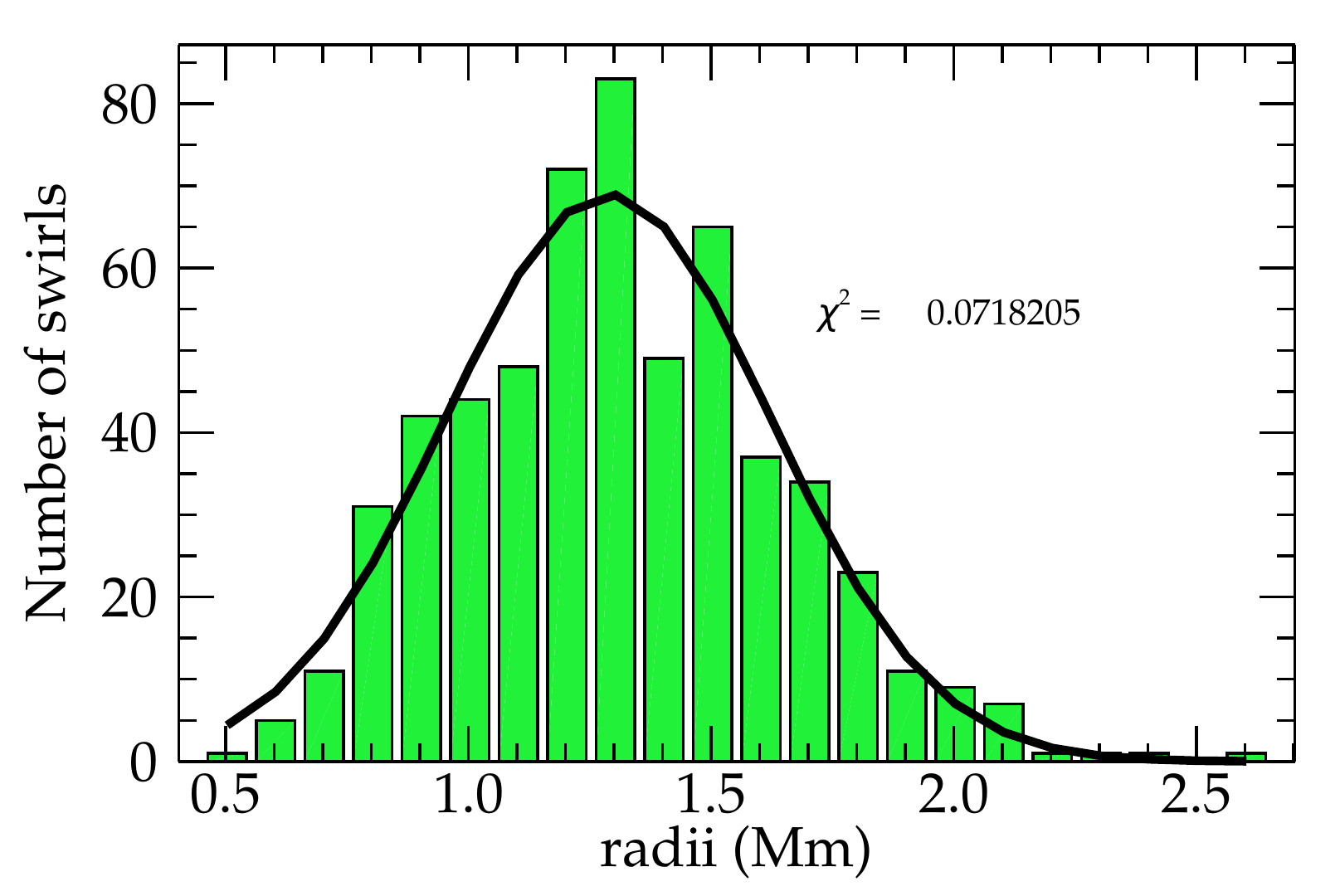}}
\caption{Histogram of the derived radii of the 577 swirls. The green bars refer to the derived radii (see text), while the overplotted black line represents the curve that fits a Gaussian distribution.}
\label{fig:mean_rad}
\end{figure}

One of the most crucial spatial parameters of swirls that can be determined is the radius. The radius of each one of the 577 swirls was calculated as the mean distance of the border segments around the swirl center. The outermost 10\% segments of the swirl were considered as border segments and the radius is calculated as:
\begin{equation}%------- EQ.1
R_{10\%}=\frac{10\cdot\sum_{j>(N_{s}/10)} d_{j}}{N_{s}},
\end{equation}\label{eq:1}\noindent
where $N_{s}$ the number of segments of the swirl, $N_{s}/10$ is a rounded integer and $d_{j}=\sqrt{(x_{j}-X_{s})^{2}+(y_{j}-Y_{s})^{2}}$ is the Euclidean distance of each segment with central coordinates $(x_{j},y_{j})$ from the swirl center $(X_{s},Y_{s})$.
\begin{figure}[h!]%--------------- FIGURE 6 - swirls segments
\resizebox{\hsize}{!}
{\includegraphics[width=0.9\textwidth]{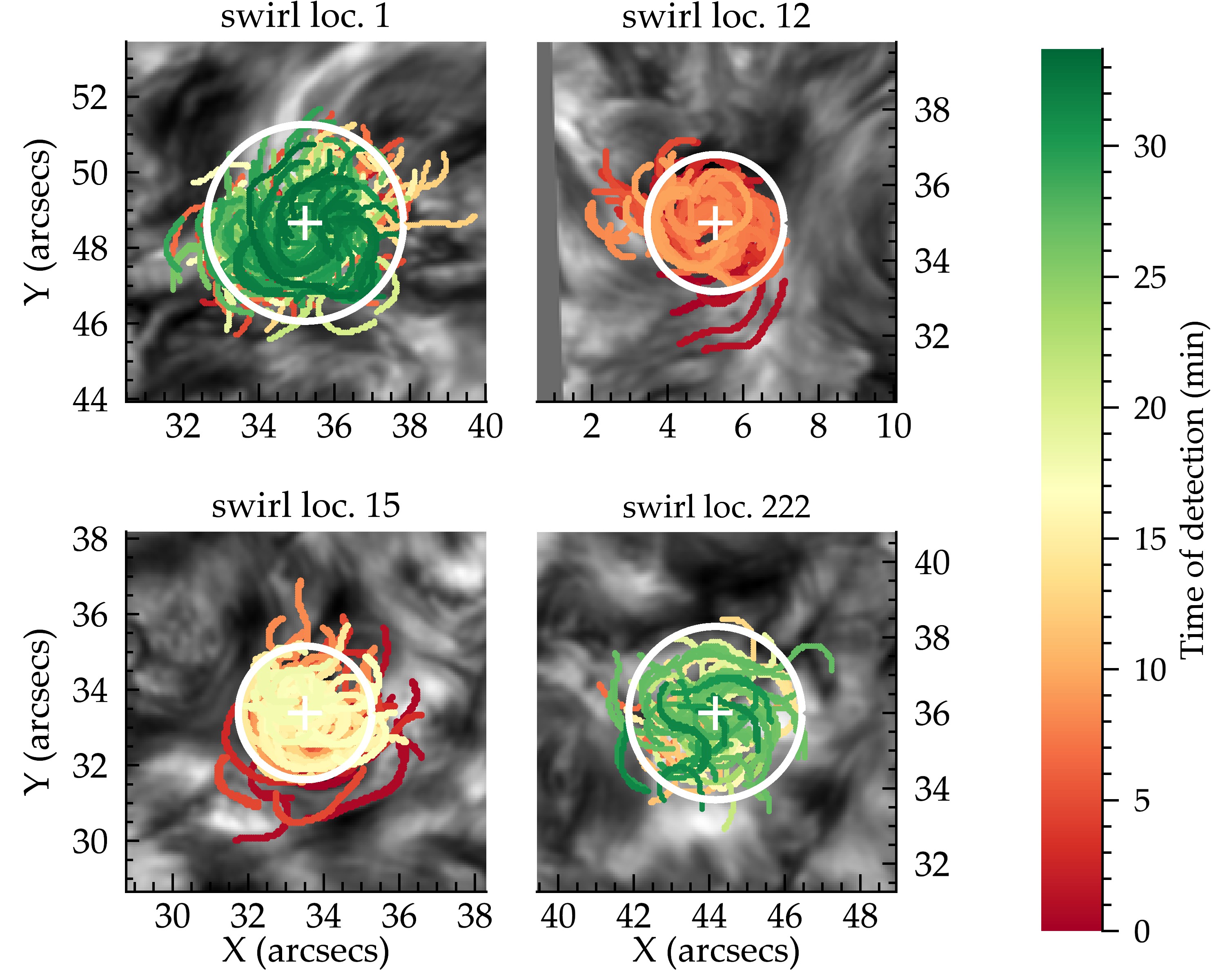}}
\caption{Segments of 4 out of 577 detected chromospheric swirls. The radii are calculated from the 10\% most distant segments with respect to the swirl center. The segments are color-coded with a colomap indicating the time of detection.  }
\label{fig:segments}
\end{figure}
Using this approach, we can avoid misleading radii derivations generated by outlier segments, while maintaining a sense of the border of the swirl, as demonstrated in the examples of Fig. \ref{fig:segments}. The resulted histogram of radii presented in Fig. \ref{fig:mean_rad} shows that they range between $\sim$0.5 to $\sim$2.5\,Mm with a mean radius of $\overline{R}=1.3\pm 0.3$\,Mm that is obtained by fitting a Gaussian density function to the histogram.

\subsection{Lifetimes of swirls}\label{subsec:lifetimes}

Lifetimes of solar events are an ambiguous physical parameter to determine, since they depend on a range of aspects, such as the duration, cadence and quality of the observations, nature, morphology, and the dynamics of the events under study. We must also take into consideration the method used for their identification and tracking, the exact definition of the term ``lifetime'' itself, and so on. In works based on visual inspection of chromospheric swirls, the lifetime is perceived as the temporal interval in which a swirl is visible in observations and presents a sense of rotation \citep{Wede12} or by the duration over which the structures can be visually tracked \citep{Shetye19}.

\begin{figure}[htb!]%--------------- FIGURE 8 - lifetime histogram
{\includegraphics[width=0.45\textwidth]{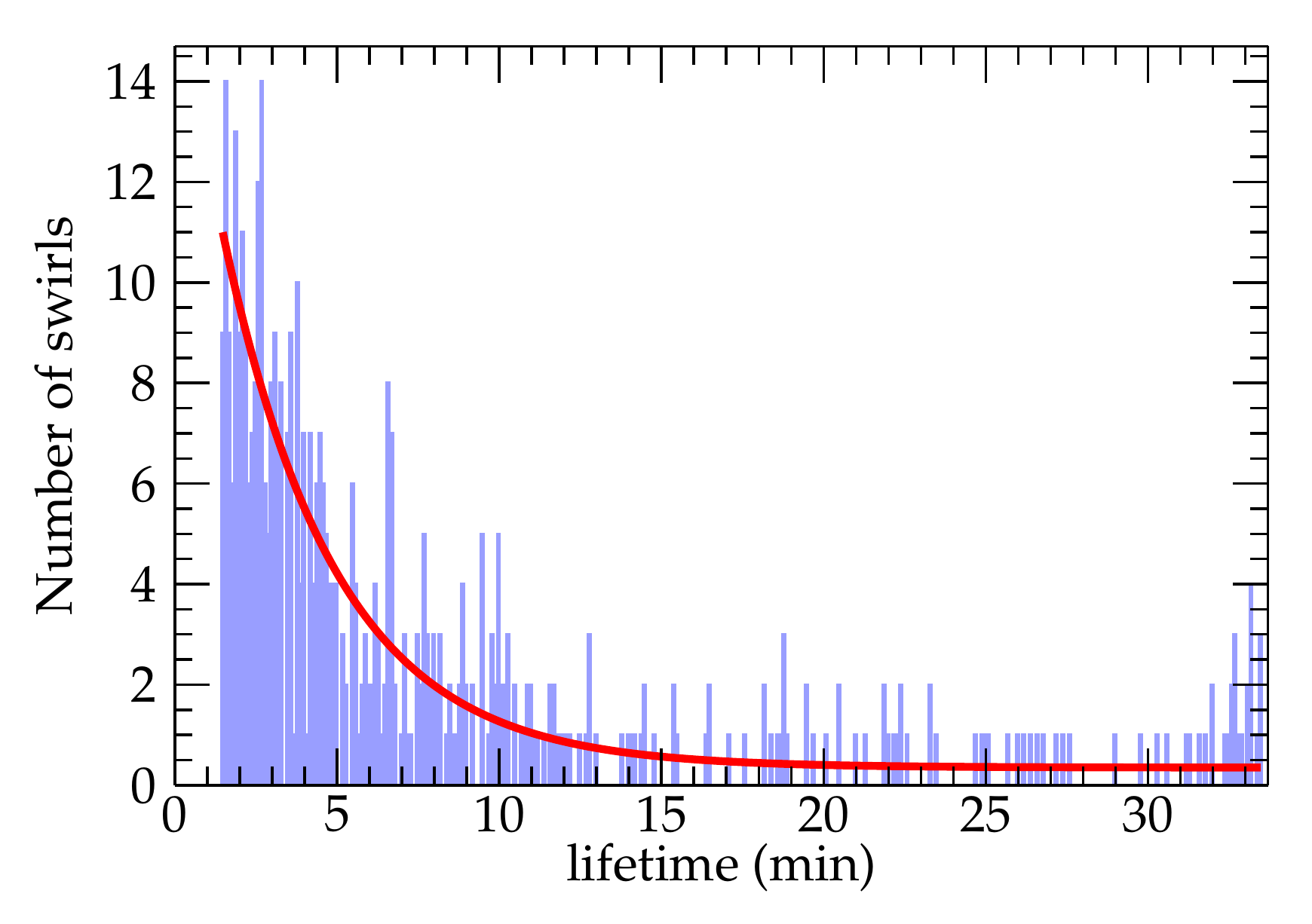}}
\caption{Histogram of the lifetimes of the 487 individual swirls within the common FoV in Fig. \ref{fig:results}. The red line represents the exponential model that fits the data.  }
\label{fig:life_histo}
\end{figure}

In the present work, after identifying and tracking the swirls in 350 locations throughout the data series, we imposed some criteria for a swirl to be considered as a distinct event (see Sec. \ref{subsec:preparation}).
The tracked events and their lifespan in each one of the 350 locations are shown in Fig. \ref{fig:lifespans}, where not only the spatial location, but also the temporal evolution of individual swirls are depicted. Due to the displacement of the FoV after 26\,min from the beginning of the observations (see Sec. \ref{sec:obs} and Fig. \ref{fig:results}), the lifetime is estimated only for those swirls that are found on the common FoV and can be tracked throughout the entire time sequence, namely, for 487 swirls. The histogram of the measured lifespans of these swirls is given in Fig. \ref{fig:life_histo}. The lifetimes range from 1.5\,min (the minimum time imposed by the algorithm) to 33.7\,min (the duration of the observations). The shorter-lived swirls, namely, those with lifetimes lower than $\sim$10\,min are more abundant. However, as Figs. \ref{fig:life_histo} and \ref{fig:lifespans} indicate, there is a considerable number of individual swirls whose lifetime is comparable or even exceeds the duration of the observations and others that are observed in the first frame or in the last frame whose existence before or their survival after the current observations, respectively, is unknown.

Performing an exponential fit of the form of $f(t)=N_{0}e^{-(t/\tau)}$ to the histogram of Fig. \ref{fig:life_histo}, we obtain a mean lifetime $\overline{T}=\tau= 3.4\pm0.2$\,min. Obviously, this value represents the mean lifetime of short-lived swirls. However, the exponential curve fails to provide a fit that matches the whole distribution of lifetimes, as it is unable to take into account the lifetimes of long-lived swirls. An arithmetic mean lifetime of the swirls of the sample can also be derived as the sum of all individual lifespans divided by their number, that is, $\overline{T}=\frac{\sum_{s}{t_{s}}}{S}\simeq8.5$ min, $s=1,..,S$, where $S=487$ is the total number of lifespans (or swirls). We note, however, that even this approach leads to an underestimation of the actual mean lifetime, which is caused by the inevitable underestimation of the contribution of swirls, whose lifetimes are comparable or exceed the total duration of the observations, or which appeared before or continue to survive after the current observation interval.

In previous works, various other approaches have been adopted in order to estimate the lifetimes of solar features, such as granules or bright points \citep{Title89,Brandt91,Utz10,Abramenko10}. One approach is to dismiss events that are observed in the first or last image and to measure lifetimes as the time difference between the first and last detection of each event \citep{Utz10}. Another approach involves the adoption of a correction factor or weight, which is used in order to counterbalance the underestimated number of long-lived features \citep{Danilovic10}, whereas in other cases, the populations of long-lived and short-lived solar features are analyzed separately and discrete mean lifetimes are extracted from the fit of their histograms (e.g., \citealp{Delmoro04}).

An extensively used method for statistical data analysis for which the outcome variable is time-to-event is the survival analysis. In a survival analysis, one feature has survived after some follow-up period, which is referred to as the "survival time." In solar physics, survival analyses were used in studies of solar granulation  to estimate the mean lifetime of granules from photographic observations (see, e.g., \citealp{Janssens70}). It consisted of selecting a well-defined number of features on a ``master'' frame at the middle of a time series of photographs and tracing them back to formation and forward to disappearance. From the survival times of the features, a survival curve is constructed from which mean lifetimes of features can be obtained \citep{Alissandra87}. Such an analysis, however, cannot be applied to the present observations because it does not take into account the lifetimes of a considerable number of swirls whose survival times are not known because of the limited duration of the observations, but also because it ultimately takes into account a small statistical sample.

Appropriate survival analysis of data requires specific statistical methods that can consider a key problem called ``censoring.'' Censoring occurs when we do not know the exact survival time of a feature; for instance, if a feature cannot be followed-up because the observations have ceased, it is referred to as "right-censored." On the other hand, a feature is referred to as "left-censored" when it appears in the first frame of the observations and, therefore, we do not know when exactly it was formed. Survival analyses that take into account censored data are widely used in several scientific fields, such as medical sciences, biological sciences, social sciences, and economics, and they are based upon survival probabilities (see, e.g., \citealp{Hosmer99}; \citealp{Klein05}; \citealp{Moore16}). The results can be significantly biased if censoring is not taken into account. This kind of analysis is very appropriate for the estimation of the mean lifetime of swirls from the present observations.

In our case, the time-to-event is the time to the appearance or disappearance of a swirl. The probability that swirls with a survival time T (which is a random variable) disappear later than some specified time, t, is given by the survival function:
\begin{equation}
S(t)=Pr(T\ge t)$, $t>t_{0},
\end{equation}
where $t_{0}$ is the initial time of the observations. $S(t)$ must be, by definition, non-negative and non-increasing with time. The probabilities over time are presented by the survival curve, which gives the proportion of the initial population that survives after time $t$. One of the assets of the survival analysis is that features in a population that have not been subject to the time-to-event (appearance and disappearance) within the observational window are labeled as left- or right-censored and are taken into consideration during the construction of the survival curve \citep{Leung97,Turk21}.
 \begin{figure}[htb!]%--------------- FIGURE 9 - survival curve
%\resizebox{\hsize}{!}
{\includegraphics[width=0.45\textwidth]{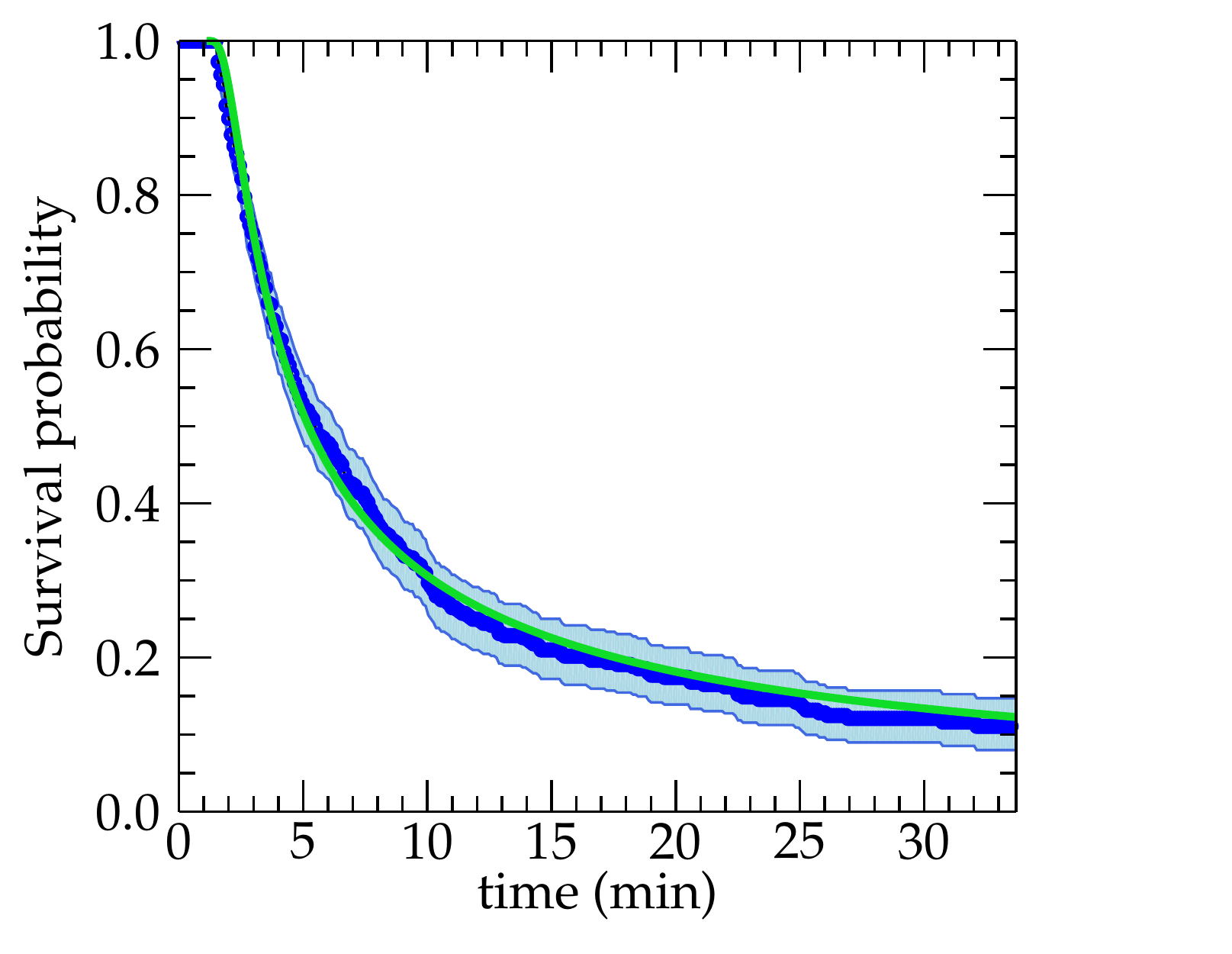}}
\caption{Survival curve constructed by the Kaplan-Meier estimator method (blue points) and the curve that fits it using the generalized gamma function (green line). The shaded blue area indicates the confidence interval of the Kaplan-Meier estimator. }
\label{fig:surv_model}
\end{figure}

There are several methods proposed for estimating the survival probability from survival times both censored and uncensored. A classic non-parametric method for estimating the survival curve taking into account both censored and uncensored lifetime data is the Kaplan-Meier Estimator \citep{Kaplan58}. This method accepts as input durations (in our case are the lifespans of swirls) described by the random variable T, and indicators of censoring, that is, right (unobserved), or no censoring (disappearance) for each lifespan in the sample in order to estimate the survival curve. Using durations as input means that the appearance of all swirls is considered to occur simultaneously, therefore, $t_{0}=0$, which implies three basic assumptions in order for this method to be valid: (a) the disappearance of the right-censored swirls is bound to occur after the end of the observations, (b) the characteristics of the observed solar region do not change drastically during the entire duration of the observations, (c) swirls whose birth occurred before the beginning of the observations can be treated as right-censored, as well (reversed left-censoring). From the resulting survival curve (presented in Fig. \ref{fig:surv_model}), we are able to easily deduce the expected lifetime of 50\% of the population, namely, the median lifetime of swirls, which is equal to 5.5\,min. The confidence intervals are calculated using the Greenwood method described in \citet{Greenwood26} and \citet{Hosmer99}. Due to the existence of a considerable number
of swirls with lifetimes equal or exceeding the duration of the observations, the survival curve does not end at zero. Finding a survival function that best fits the survival curve enables the calculation of significant lifetime-correlated parameters that cannot be calculated otherwise, since the Kaplan-Meier estimate does not provide an analytical model. Several survival models that are associated with well-known probability density functions and are frequently used in lifetime data analyses have been applied, such as the Gaussian, the Weibull, and the lognormal distributions. However, in order to determine the most appropriate parametric model we used generalized gamma distribution $f(t;\mu,\sigma, \lambda)$, which incorporates the exponential, Weibull, and lognormal as sub-models, depending on the values of the shape parameters $\mu,\sigma, \lambda$. The survival function takes the parameterized form:
$$S(t)=\begin{cases}%------- EQ.2
1-\Gamma_{RL}\left( \frac{1}{\lambda^2};\frac{e^{\lambda\left(\frac{log(t)-\mu}{\sigma}\right)}}{\lambda^{2}} \right), \quad \mbox{if    } \lambda > 0 \nonumber \\
\Gamma_{RL}\left( \frac{1}{\lambda^2};\frac{e^{\lambda\left(\frac{log(t)-\mu}{\sigma}\right)}}{\lambda^{2}} \right), \quad \mbox{if    } \lambda \le 0,
\end{cases}\label{eq:2}
$$
where $\Gamma_{RL}$ is the regularized lower incomplete gamma function. The fit of the model resulted to the parameters $\lambda=-2.36$, $\mu=3.27$ and $\sigma=0.56552547$ derived by using standard fitting routines of the Python lifelines library \citep{David19}. Based on the selected model, we calculate the mean lifetime as:
\begin{equation}%------- EQ.3
\overline{T}=\int_{0}^{t_{f}} S(t)\,dt,
\end{equation}\label{eq:3}\noindent
where $t_{f}$ is the time referring to the end of the observations.
These calculations yield $\overline{T}=10.3\pm0.6$\,min. The theoretical survival function becomes by definition asymptotic to zero, while the integral of the mean lifetime is calculated from zero to infinity. This indicates that our result is still delivering an underestimation of the actual mean lifetime. However, this result approximates more accurately than other methods the latter, because it encompasses both cases of swirls: those with censored lifetimes and longer-lived swirls, which altogether represent a considerable proportion of our sample ($\sim$20\%). An expansion of the model in order to extrapolate the survival function beyond the observational window is possible, but it is not safe to assume the evolution and behavior of swirls without factoring in the contribution from the observations.

\subsection{Correlation between lifetimes and radii of swirls}

\begin{figure}[htb!]%--------------- FIGURE 10 radii - lifetimes
%\resizebox{\hsize}{!}
{\includegraphics[width=0.40\textwidth]{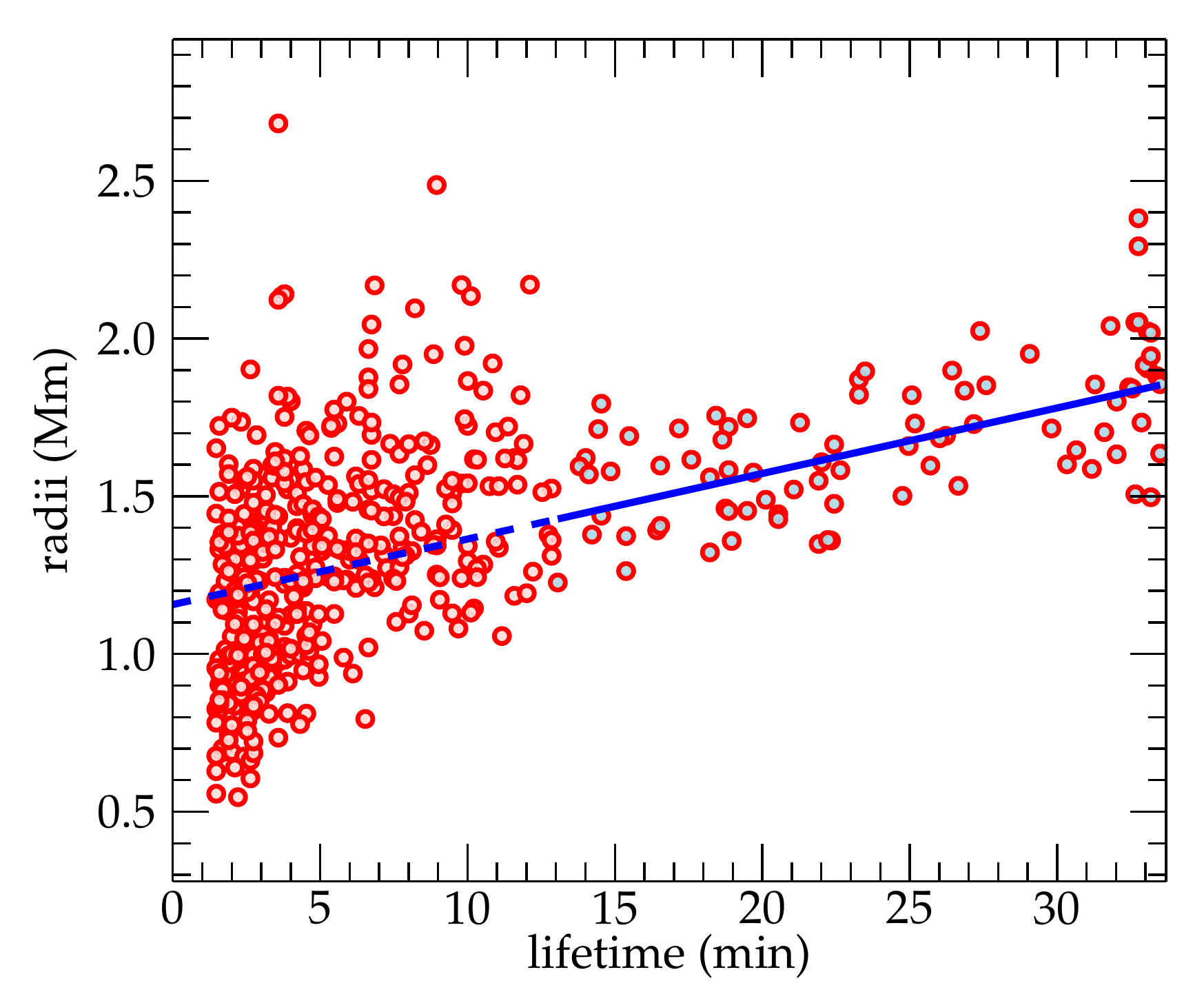}}
\caption{Scatter plot of the radii of the swirls (see text) vs. their lifetimes. The linear fit was calculated for swirls with a lifetime larger than 13 min (blue line) with a Pearson coefficient equal to 0.63. }
\label{fig:rad_life}
\end{figure}

A scatter plot between the obtained radii of the 487 swirls that reside within the common FoV and their lifetimes is shown in Fig. \ref{fig:rad_life}. Swirls with lifetimes lower than $\sim$13\,min have a wide range of radii extending from 0.5 to 2.2\,Mm with no obvious correlation. In contrast, radii of longer-lived swirls seem to show positive linear correlation with lifetime. The Pearson coefficient for swirls with lifetimes greater than 13 min was found equal to 0.63 and a linear least-squares fit yields:
\begin{equation} %-----EQ4
T=0.02R_{10\%}+1.16,
\end{equation}\noindent
where $T$ is the random variable of lifetimes greater than 13 min and $R_{10\%}$ are the radii of the swirls as they were calculated in Sec. \ref{subsec:radii}. Such a correlation was not found before in observations or simulations. While \citet{Kato17}   did not find an obvious relation between lifetime and size in their simulation-based sample, they suggested there is a need for a more systematic analysis. Swirls with lifetimes higher than 13\,min have radii larger than 1.2\,Mm.

\subsection{Spatial density and occurrence rate of swirls}

After applying the restrictions mentioned in the previous sections, a mean number of $\overline{N}_{s}=\frac{\sum_i N_i}{N}=146\pm 9$ swirls is found in the H$\alpha$-0.2\,{\AA} FoV at any given time, where $N_{i}=N(t_{i})$ is the number of swirls at each time step, $t_{i}=t_{0}+cadence\cdot i$, $t_{0}$ is the initial time of observations, and $N$ is the number of images with $Q_{f}>0.4$. The mean surface density is calculated by dividing the mean number of swirls, $\overline{N}_{s}$ by the FoV area, namely by 1737\,Mm$^{2}$, which yields $\sim$0.08\,swirls\,Mm$^{-2}$. The occurrence rate is obtained by dividing the 487 detected swirls in the common FoV of 1447\,Mm$^2$ by both the FoV area and the duration of the current observations and is equal to $\sim$10$^{-2}$\,swirls\,Mm$^{-2}$\,min$^{-1}$. Finally, by dividing the mean surface density by the occurrence rate, we derive an additional approximation of the mean lifetime of the chromospheric swirls, which is $\sim$8.5\,minutes. This is in agreement with the arithmetic mean lifetime, which was calculated directly by the durations of the individual swirls in Sect. \ref{subsec:lifetimes}. From the mean radius, which is equal to 1.3\,Mm, the area covered by chromospheric swirls at any given time is $\pi\overline{R}^{2}\overline{N}_{s}\simeq 775$\,Mm$^{2}$ in an overall 1927\,Mm$^{2}$ quiet-Sun area, where the additional area due to the displacement of the pointing is included. Consequently, in the current observations, approximately 40\% of the entire FoV is covered by chromospheric swirls at any given time.

\section{Summary and discussion}\label{sec:disc}

In the present work, we use the novel automated detection method introduced in Paper I to detect and track chromospheric swirls and to perform a statistical analysis of their parameters. This method is capable of detecting and tracking curved structures, such as chromospheric swirls, based on their morphological characteristics. In Paper I, the method was applied to chromospheric observations also obtained in the H$\alpha$ line with the SST/CRISP instrument, but with a shorter 6.5\,min duration. The method proved to be very efficient in identifying and tracking chromospheric swirls in both space and time within the examined FoV. In this work, the method is applied to a new high spatial and temporal resolution H$\alpha$ time series of $\sim$34\,min duration obtained again with the SST/CRISP instrument.

The region we observed was very quiet and was found almost at the solar disk center. The high quality of the data due to the high capabilities of the instrument, the high cadence and the good seeing conditions, for most of the time, revealed the unambiguous presence of a large number of swirls within the observed FoV which is obvious even through visual inspection. Such a large number of swirls has never been seen thus far in chromospheric observations. The algorithm was applied to the H$\alpha$-0.2\,{\AA} dataset where the swirls are more visible than in the H$\alpha$ line center, and its performance was carefully investigated and tested. The same parametrization as the one presented in Paper I (with minor modifications) was used for the automated detection and tracking of swirls. The results were analyzed to obtain the statistical distributions of their physical parameters, such as radii and lifetimes, as well as their abundance and occurrence rate.
\begin{table*} %----------------TABLE
\caption{Properties of photospheric vortices and chromospheric swirls derived via observations and simulations. }
\label{tab:1}
\centering
\begin{tabular}{p{0.25\linewidth} c c c c c}
\hline\hline
 & space density  & occurrence rate  & radius  & lifetime \\
 & (vortices\,Mm$^{-2}$) & (vortices\,Mm$^{-2}$\,min$^{-1}$) & (Mm) & (min)  \\ %--table heading
\hline
\multicolumn{5}{c}{Photospheric observations}\\
\hline
\citet{Brandt88} &  & & 0.25 & > 90  \\
\citet{Bonet08} & 9 $\times 10^{-3}$ & 1.8$\times 10^{-3}$  & $\la 0.5$ & 5.1\,$\pm$\,2.1 \\
\citet{Attie09} &  & & $\sim$7.5 \& $\sim$10.5 & >90 \& >120 \\
\citet{Bonet10} & 2.4$\times 10^{-2}$  & 3.1$\times 10^{-3}$ & 0.5 & 7.9\,$\pm$\,3.2 \\
\citet{Balmaceda10} &  &  & 0.9 & $\ga 20$ \\
\citet{Vargas11} & 2.8$\times 10^{-2}$ & $\sim$1.4$\times 10^{-3}$ & 0.25 & $\sim$10 - 20 \\
\citet{Reque18} & & & 2.5 & $\sim$420 (7 h)  \\
\citet{Giagkio17} & 0.24 & 0.84  & $\sim$0.28 & 0.29   \\
\citet{Liu19} & 6.1$\times 10^{-2}$ & & 0.29\,$\pm$\,0.064 & 0.35  \\
\hline
\multicolumn{5}{c}{Chromospheric observations}\\
\hline
\citet{Wede09} & &  1.24$\times 10^{-4}$  & 0.75  & \\
\citet{Wede12} & 2$\times 10^{-3}$ &    & 1.5\,$\pm$\,0.5   & 12.7\,$\pm$\,4 \\
\citet{Park16} &  & & 0.3 \& 0.7 & 1 \& 2 \\
\citet{Tzio18} &   & & $\sim$2.2 & >102  \\
\citet{Shetye19} & & & $\sim$1 & $\sim$10 \\
Present study & $\sim$8$\times 10^{-2}$ & $\sim$10$^{-2}$  & 1.3\,$\pm$\,0.3 & 10.3\,$\pm$\,0.3 \\
\hline
\end{tabular}
\end{table*}

The application of the algorithm resulted to the detection of highly curved segments in each image of the dataset. Using a first-level clustering in space the highly curved segments are grouped together, while a second level clustering labels as swirls the more persistent ones. In addition, it provides their centers and area, and their evolution in time. There are temporal gaps in the detection of swirl centers that appear at the same locations that are due to missing frames or low-quality images, or to the dynamic nature of the chromosphere, whereby the algorithm is unable to trace a curved segment, or to some underlying physical process that causes a structure to disappear and reappear at the same location. To tackle this issue, an additional processing step was adopted. This step consists of filling-in temporal gaps and considering as belonging to the same event structures whose centers are found at the same location and have gaps in their temporal evolution that are less than or equal to a calculated threshold. In addition, 11 detected curved structures, found to correspond to elongated fibrils that were slightly curved at one of their edges, were excluded from the sample.

The derived sample consists of 577 swirls in 350 swirl locations within the examined FoV of 1737\,Mm$^{2}$. Due to a slight displacement of the pointing during the last minutes of the observations, the total FoV covered from the two pointings is equal to 1927\,Mm$^{2}$, while within the common FoV, which is equal to 1447\,Mm$^{2}$, 487 swirls out of the 577 can be followed in time during the entire time sequence and have been used for the temporal statistical analysis. A mean number of 146\,$\pm$\,9 swirls are detected within the H$\alpha$-0.2\,{\AA} FoV at any given time. The mean surface density is $\sim$0.08\,swirls\,Mm$^{-2}$ and the occurrence rate is equal to $10^{-2}$\,swirls\,Mm$^{-2}$\,min$^{-1}$. These numbers are much higher than those reported before from chromospheric observations. The occurrence rate is two orders of magnitude higher compared to the estimation of \citet{Wede09} who found 10 clear examples of swirls, 10 less clear, and 16 potential but less reliable detections within a FoV of $71\arcsec \times$ 71$\arcsec,$   giving a swirl occurrence rate of 1.24$\times 10^{-4}$\,swirls\,Mm$^{-2}$\,min$^{-1}$. \citet{Wede12} identified 14 swirls within a $55\arcsec \times$ 55$\arcsec$ FoV, a space density 2$\times 10^{-3}$\,swirls\,Mm$^{-2}$, and an occurrence rate of 1.9$\times 10^{-4}$\,swirls\,Mm$^{-2}$\,min$^{-1}$. In both works, the swirls were identified by visual inspection in \ion{Ca}{II}\,8542\,{\AA} SST/CRISP observations, while the region observed in the former work was in a coronal hole. In Paper I, when applying the automated code to an 6.5\,min H$\alpha$ SST/CRISP time series, 10 swirls were detected within a $60\arcsec \times$ 60$\arcsec$ FoV. We note that a large portion of the investigated FoVs in the work of \citet{Wede12} and in Paper I is covered by a high number of linear fibrils, which seem to expand above areas with a high concentration of photospheric magnetic elements. These areas are obviously void of swirls. In the FoV of the present dataset, the portion covered by linear fibrillar structures is lower, allowing the detection of a higher swirl population. In Table \ref{tab:1} different parameters derived so far from photospheric and chromospheric observations are summarized along with the results of the present study.

One advantage of the automated code is that the centers of individual swirls and the curved segments can be traced in each image of the time series. The radii of the detected swirls can be estimated from the outer traced segments. They have a Gaussian distribution and range between 0.5 and 2.5\,Mm with a mean value equal to 1.3\,$\pm$\,0.3\,Mm. These values are slightly higher than the values previously reported in literature for chromospheric observations, but substantially higher than those derived from photospheric observations or from simulations (see the introduction). The Gaussian distribution and the lower estimated value of the radii means that the lower imposed value in the algorithm of 0.34\,Mm for the radius of a swirl is a reliable value.

The lifetime of swirls is an important parameter to determine, since it is essential for their modeling and their energetics. Its determination, however, is a challenging task because it is highly dependent on the observational dataset, the method used for tracking the features under study, the assumptions made and the definition of the term itself. In the present work, the lifetimes of the detected swirls in the common FoV are estimated as the time difference between the first and last frame of their appearance and disappearance, respectively, during the entire duration of the observations. The values of lifetimes range from 1.5\,min (minimum value imposed by the algorithm) to $\sim$33.7\,min (duration of the observations). The form of their distribution may imply the existence of two populations of swirls, but for this suggestion to be confirmed, a longer time series is needed. From this distribution, it is found that most of the swirls have lifetimes lower than 13\,min. There is, however, a considerable number of features that have lifetimes comparable to the duration of the observations, as well as others that are observed in the first or in the last frame for which we do not know if they existed before the beginning of the observations or continue to exist after the end of the observations. These cases combined represent $\sim$20\% of the sample. Taking the above into account,  the arithmetic mean lifetime derived from all swirls in the sample without rejecting any of them is found $\sim$8.5\,min, which is clearly an underestimation of their actual duration. An exponential fit to the distribution of the lifetimes gives a mean lifetime of 3.5\,$\pm$\,0.3\,min. Obviously, this value represents the mean lifetime of the short-lived swirls because the exponential function is unable to represent the entire lifetime distribution which includes both short- and long-lived swirls.

A statistical analysis of lifetimes that is relevant to the present data for which a proportion of survival times are unknown is the survival analysis. The survival function represents the probability that an event survives from the time of origin to some time beyond time, $t,$ and can be estimated through various methods. In this work, we use the Kaplan-Meier method and we constructed the survival curve. It is found that an analytical model that sufficiently represents the survival curve is the generalized gamma function. Based on this model, the mean lifetime of swirls is found to be equal to 10.3\,$\pm$\,0.6\,min. This is a more reliable value for the mean lifetime, but it is still an underestimation because of the limited duration of the present observations. The value of the mean lifetime obtained by the survival analysis is close to the values obtained for a low number of swirls from chromospheric observations by visual inspection (see  the introduction), but much higher than the values reported from photospheric observations or simulations. It should be noted, however, that there are swirls lasting for the entire duration of chromospheric observations, namely, at least 1.7\,h \citep{Tzio18}.

A correlation analysis between lifetimes and radii shows that longer-lived swirls tend to have larger radii with a positive linear relationship between these parameters, while shorter-lived swirls are more abundant and have smaller, scattered radii without a clear relationship between lifetimes and radii. This behavior may point to the fact that larger swirls last longer than the smaller ones.

The novel automated method developed, presented, and implemented in Paper I,  and implemented in the present work as well, is the only method proposed so far that is based on the morphological characteristics of chromospheric swirls and capable of identifying and tracking them in space and time in observational datasets. Applying it in time series allows for the evaluation of some of their statistical properties. In the
current literature, there are only a few statistics that are  based on chromospheric swirl detections from such a large sample as the one used in the present work. One aspect that warrants mention is that we found that $\sim$40\% of the FoV is covered by swirls at any given time. This is a compelling difference compared with previous observations. The reason for the large number of swirls detected in the current observations remains to be found. As a suggestion it may be due to: a) the application of an automated code, b) the quality of the data, or c) the observed solar region itself, which although it is not in a coronal hole, it is very quiet and the physical properties of the underlying photosphere together with the magnetic field topology may favor the formation of swirls in the chromosphere.

It should be mentioned that the accuracy of every automated detection and tracking method has some limitations. Its efficiency depends on the quality and the temporal and spatial resolution of the available dataset. The requirements imposed by the algorithm for the minimum radius and lifetime may also affect the results, especially the lifetime. For instance, the imposed minimum radius of 340\,km and minimum lifetime of 1.5\,min requirements for a curved structure to be traced and considered as a swirl may lead to overestimations of the mean radius and lifetime. An additional requirement we use in our analysis is a gap-filling procedure in order to overcome the effect of their disappearance between frames. This procedure could lead to lower or higher number of detected swirls and to higher or lower mean lifetimes, respectively, depending on the number of frames used. On the other hand, imposing these requirements helps to reduce false-positive detections from close to noise fluctuations.

\section{Conclusions}\label{sec:concl}

Vortex flows observed in the solar photosphere are fundamental indicators of the turbulence due to convective flows. The interplay between these flows and the magnetic field forms twisting magnetic field lines that penetrate the solar layers up to the low corona, leading to some interesting dynamical phenomena, such as chromospheric swirls and magnetic tornadoes. Successful detections and tracking of swirls with an automated method applied to good-quality, high-cadence, and sufficiently long observational datasets is a key step toward quantifying their physical and statistical properties. Our results show they are very abundant in the chromosphere. As these structures are considered to play an important role in the generation of MHD waves \citep{Erdel06,Jess09,Fedun11,Shel13,Tzio20}, the propagation of energy, and the heating of the solar outer atmosphere \citep{Wede12}, accurate determinations of their abundance and  properties are necessary for the quantitative determination of their contribution.

An important step to improving the understanding of chromospheric swirls is to apply the automated detection code to other chromospheric lines, such as  \ion{Ca}{II}\,IR or \ion{Ca}{II}\,H, and perform a similar statistical analysis. The automated detection code as presented in Paper I and used in this work is designed in such a way that the key parameters can be easily modified, and the code can be applied to other observational data and in other chromospheric lines in which these features are observed. In addition, a multiwavelength analysis of the detected swirls in various spectral lines will permit the derivation of their physical parameters in various chromospheric or transition region heights.

Studies through visual inspection or from the detection of chromospheric swirls in simulations using local correlation tracking methods have shown (although not definitively) that chromospheric swirls may be associated with photospheric magnetic concentrations located in the intergranular lanes. A comparative study of the locations of photospheric magnetic concentrations and chromospheric swirls may allow for establishing their association, along with identifying photospheric-chromospheric-coronal  coupling and multi-layer characteristics of swirling events.

Pushing the limits of observations to smaller scales, namely, at higher cadences and higher spatial resolutions, will most probably result in the detection of swirls that have gone undetected with current telescopes. New powerful solar telescopes, such as DKIST or the upcoming EST, are expected to provide clearer observational evidence about their abundance and the range of their physical parameters. In this context advanced techniques of automated detection and analysis should be considered as a prerequisite, while existing methods for the estimation of the velocity field require certain improvements, such as the inclusion of machine learning techniques \citep{Asensio17}, in order to assure that they are up to the task.

\begin{acknowledgements}
This research leading to the results obtained has been supported by the SOLARNET project that has received funding from the European Union's Horizon 2020 research and innovation programme under grant agreement No.824135, the PRE-EST project under grant agreement No. 739500 and by national funding. We are indebted to G. Vissers for handling the SST service mode observations, the coordination with IRIS co-observations and the reduction of the SST data obtained during this observational campaign. The Swedish 1-m Solar Telescope is operated on the island of La Palma by the Institute for Solar Physics of Stockholm University in the Spanish Observatorio del Roque de los Muchachos of the Instituto de Astrofisica de Canarias. This work is part of a collaboration between the National Observatory of Athens and AIP supported by IKYDA2020, an action program funded by the German Academic Exchange Service (DAAD), under project No.57546881, and the Greek State Scholarship Foundation (IKY). The authors would also like to thank the International Space Science Institute (ISSI) in Bern, Switzerland, for the hospitality provided to the members of the team on ``The Nature and Physics of Vortex Flows in Solar Plasmas''. Finally, we thank the referee, Viktor Fedun, for the useful and insightful notes and comments, which helped to improve the paper.

\end{acknowledgements}

\bibliographystyle{aa}
\bibliography{stats_biblio}

\end{document}